\documentclass[article]{jss}
\pdfoutput=1
\makeatletter
\def\maxwidth{ %
  \ifdim\Gin@nat@width>\linewidth
    \linewidth
  \else
    \Gin@nat@width
  \fi
}
\makeatother

\definecolor{fgcolor}{rgb}{0.251, 0.251, 0.282}
\newcommand{\hlnum}[1]{\textcolor[rgb]{0.125,0.125,1}{#1}}%
\newcommand{\hlstr}[1]{\textcolor[rgb]{0.125,0.125,1}{#1}}%
\newcommand{\hlcom}[1]{\textcolor[rgb]{1,0,0.753}{\textit{#1}}}%
\newcommand{\hlopt}[1]{\textcolor[rgb]{0.251,0.251,0.282}{#1}}%
\newcommand{\hlstd}[1]{\textcolor[rgb]{0.251,0.251,0.282}{#1}}%
\newcommand{\hlkwa}[1]{\textcolor[rgb]{0,0.533,0.345}{\textbf{#1}}}%
\newcommand{\hlkwb}[1]{\textcolor[rgb]{0.439,0.251,1}{\textbf{#1}}}%
\newcommand{\hlkwc}[1]{\textcolor[rgb]{0.529,0,0.184}{\textbf{#1}}}%
\newcommand{\hlkwd}[1]{\textcolor[rgb]{0.251,0.251,0.282}{\textbf{#1}}}%

\usepackage{framed}
\makeatletter
\newenvironment{kframe}{%
 \def\at@end@of@kframe{}%
 \ifinner\ifhmode%
  \def\at@end@of@kframe{\end{minipage}}%
  \begin{minipage}{\columnwidth}%
 \fi\fi%
 \def\FrameCommand##1{\hskip\@totalleftmargin \hskip-\fboxsep
 \colorbox{shadecolor}{##1}\hskip-\fboxsep
     \hskip-\linewidth \hskip-\@totalleftmargin \hskip\columnwidth}%
 \MakeFramed {\advance\hsize-\width
   \@totalleftmargin\z@ \linewidth\hsize
   \@setminipage}}%
 {\par\unskip\endMakeFramed%
 \at@end@of@kframe}
\makeatother

\definecolor{shadecolor}{rgb}{.97, .97, .97}
\definecolor{messagecolor}{rgb}{0, 0, 0}
\definecolor{warningcolor}{rgb}{1, 0, 1}
\definecolor{errorcolor}{rgb}{1, 0, 0}
\newenvironment{knitrout}{}{} % an empty environment to be redefined in TeX

\usepackage{alltt}

\usepackage[english]{babel} % for multilingual support
\usepackage[utf8]{inputenc} % input encoding. some others complain of frequent conflicts. i have not had many issues
\usepackage{lmodern} % latin modern fonts
\usepackage{microtype} % conserves space, and makes text prettier by making micro adjustments to text
\usepackage{amsmath} % needed for math
\usepackage{amssymb} % for bold math. see here: http://tex.stackexchange.com/a/99286
\usepackage{algorithm}  % needed for algpseudocode
\usepackage{algpseudocode}

\usepackage[matstyle=bbold]{MattsMacros}
\usepackage{listings}
\newcommand{\ourpkg}{\pkg{RefManageR}}
\newcommand{\bt}{\`{}}

\lstdefinestyle{error}{
  language=C,
  emptylines=1,
  breaklines=true,
  basicstyle=\ttfamily\color{red},
  moredelim=**[is][\color{red}]{@}{@},
}

\lstdefinestyle{output}{
  language=C,
  emptylines=1,
  breaklines=true,
  basicstyle=\ttfamily\color{black},
}

\title{Straightforward Bibliography Managament in \R{} Using the \ourpkg{} Package}

\date{\today}

\author{Mathew W.\ McLean\\ Texas A\&M University}
\Address{Mathew W.\ McLean\\ 
Institute for Applied Mathematics and Computational Science\\
Texas A\&M University\\
3143 TAMU\\
College Station, TX, 77843\\  
E-mail: \email{mmclean@stat.tamu.edu}\\
URL: \url{http://stat.tamu.edu/~mmclean}
}

\Abstract{This work introduces the \proglang{R} package \ourpkg{}, which provides tools for importing and working with bibliographic references.  It extends the \texttt{bibentry} class in \R{} in a number of useful ways, including providing \R{} with previously unavailable support for \Biblatex{}.  \Biblatex{} provides a superset of the functionality of \Bibtex, including full Unicode support, no memory limitations, additional fields and entry types, and more sophisticated sorting of references.  \ourpkg{} provides functions for citing and generating a bibliography with hyperlinks for documents prepared with \code{RMarkdown} or \code{RHTML}.  Existing \code{.bib} files can be read into \R{} and converted from \Bibtex{} to \Biblatex{} and vice versa.  References can also be imported via queries to NCBI's Entrez, \proglang{Zotero} libraries, Google Scholar, and CrossRef.  Additionally, references can be created by reading PDFs stored on the user's machine with the help of Poppler.  Entries stored in the reference manager can be easily searched by any field, by date ranges, and by various formats for name lists (author by last names, translator by full names, etc.). Entries can also be updated, combined, sorted, printed in a number of styles, and exported.  
}

\Plainkeywords{R, Biblatex, Bibtex, reference management, document generation, Unicode, cURL}
 \usepackage[hyphenbreaks]{breakurl}
\begin{document}
%\SweaveOpts{concordance=TRUE}
\maketitle
\section{Introduction}
Creating, managing, and processing references can often be a hastle.  There are a number of reasons one may want or need to work with bibliographic data in \R{} \citep{R}, for example for bibliometrics.  The \code{person} and \code{bibentry} classes available in the base-priority \pkg{utils} package since \R{} 2.14.0 provide very useful functionality for working with names and bibliographic information, respectively.  An introduction to these classes is available in \citet{hornik2012who}.  In this paper, I introduce the \ourpkg{} package, which uses these classes as building blocks to greatly simplify working with bibliographies in \R{}.

%Mention \pkg{bibtex}, \pkg{CITAN}, \pkg{scholar}, \pkg{tm}, ROpenSCi, \pkg{knitcitations}, \pkg{RMendeley}.
The \code{bibentry} class is designed to work with references in \Bibtex{} format \citep{bibtex}.  \ourpkg{} provides the \code{BibEntry} class which also works with \Bibtex{} references, but additionally supports \Biblatex{} formatting.  

The \Bibtex{} fields stored in a \code{bibentry} object can be easily accessed using the \code{\bt$\bt} operator, but there do not exist functions for conveniently conducting complicated searches.  These are provided by the \ourpkg{} package using the \code{\bt[\bt} operator.  With this operator one may search a collection of references by any field or group of fields.  \Biblatex{} fields for lists of names, such as 'author' and 'editor', can be searched by family name only, full name, or full name with initials.  Additionally, dates may be specified by ranges and are compared using the lubridate package \citep{lubridate}.  Entries may also be indexed by key, created in several different ways using functions \code{BibEntry} and \code{as.BibEntry}, and updated using the \code{\bt[<-\bt} operator.  The \code{bibentry} class provides a method for the \code{c} generic for concatenating entries, our package retains this feature, while also providing a \code{merge} method to remove potential duplicate entries when combining entries from various sources.

Entries may be imported into \R{} in a number of ways.  A function is provided for reading in \code{.bib} files in \Biblatex{} and \Bibtex{} format.  For machines with \code{Poppler} (\url{http://poppler.freedesktop.org}) installed, bibliographic metadata can be read from PDFs stored on the user's machine to generate a citation for each PDF.  The package also contributes interfaces to the CrossRef, Zotero, and NCBI's Entrez APIs to search and import references from these resources, using the \pkg{RCurl} package \citep{rcurl} for the HTTP requests.  References can additionally be obtained from a researcher's Google Scholar profile.

The package is equiped with additional printing formats and several bibliography and citation styles.  All the bibliography sorting options available in \Biblatex{} are available in \ourpkg{}.  A convenient interface for setting optional arguments for the most commonly used functions similar to the \code{options} function is used.  In case it is necessary to convert between formats, for example when submitting to a journal that does not support \Biblatex{}, a function is provided for converting a bibliography with \Biblatex{} formatting back to \Bibtex{}.  

To our knowledge our package is the first of its kind to provide support for including citations and bibliographys with hyperlinks in \code{[R]HTML} and \code{[R]Markdown} documents.  Links can point from each citation to their bibliography entry and vice versa, and hyperlinks are also automatically created for values in the \Biblatex{} fields `url', `doi', and `eprint'.

The rest of the document proceeds as follows: In Section~\ref{sec_create} I show how to create bibliography entries in \R{}, import them from local files, and discuss setting package options; Section~\ref{sec_import} discusses importing references from the web; in Section~\ref{sec_print} I discuss printing, sorting, and exporting references; in Section~\ref{sec_manip} I show how to search and update \code{BibEntry} objects; Section~\ref{sec_cite} introduces using \ourpkg{} to cite references and print a bibliography of only cited references; lastly, Section~\ref{sec_conc} concludes.
% One advantage of \Biblatex{} over \Bibtex{} is that it does not use \bst{} files for styling the bibliography. A \bst{} file must be written in a special-purpose language that few are familiar with.  One style is implemented in base \proglang{R}, namely the one used by the Journal of Statistical Software.  The code may be viewed in \proglang{R} by entering \code{as.list(tools:::makeJSS())} at the console.  On the other hand \Biblatex{} bibliographies, are styled entirely using \TeX{} macros.  Multilanguage support.  Better support for crossreferences.  No memory issues for large databases like with \Bibtex{}.  Sorting and encoding issues (discussed in biblatex doc section 2.4.3). Mention eTeX and url handling
% 
% Another advantage of \Biblatex{} is its support for UTF-8 encoding. ``bibtex is an 8bit engine so it processes every file in 8-bit pieces. In utf8 non-ascii chars are longer than 8 bit so they are splitted by bibtex. This means that bibtex has problems to sort references with non-ascii chars correctly. It can also happen that bibtex inserts a line break in the middle of an utf8-char and then you will get errors.''  A list of the supported encodings on your system can be viewed in \proglang{R} using \code{iconvlist()}.  By default, ReadBib will read in a .bib file using UTF-8.  UTF-8 is  
\section[Creating BibEntry Objects and Importing From Files]{Creating \code{BibEntry} Objects and Importing From Files}\label{sec_create}

\subsection[The BibEntry Function]{The \code{BibEntry} Function}
Similar to the \code{bibentry} function in \pkg{utils}, \ourpkg{} provides a function \code{BibEntry} for creating a \code{BibEntry} object containing a single reference, which can be combined with other references into a single \code{BibEntry} object.  An entry is specified to the \code{BibEntry} function via an argument \code{bibtype} for the entry type, an argument \code{key} for the entry key and by specifying other arguments
in \code{field = value} form to the \code{"..."} argument.  Though the `year' field is still supported in \Biblatex{} to allow backwards compatibility with \Bibtex{}, the field `date' is preferred and allows for a number of different formats for the date, which will be discussed later.  The field `journaltitle' is preferred for specifying journals, though `journal' remains supported.  Below I create and print an entry of type `Article' with fields `author', `title', `date', `journaltitle', `volume', and `number'.  The \code{print} function for \code{BibEntry} objects offers a number of features which will be discussed in detail later.  Its default settings are chosen to mimic the defaults of \Biblatex{}.  The \code{toBiblatex} function can be used to display the entry in its \code{.bib} file format.
\begin{knitrout}
\definecolor{shadecolor}{rgb}{0.973, 0.973, 0.973}\color{fgcolor}\begin{kframe}
\begin{alltt}
\hlstd{bib} \hlkwb{<-} \hlkwd{BibEntry}\hlstd{(}\hlkwc{bibtype}\hlstd{=}\hlstr{"Article"}\hlstd{,} \hlkwc{key} \hlstd{=} \hlstr{"barry1996"}\hlstd{,} \hlkwc{date} \hlstd{=} \hlstr{"1996-08"}\hlstd{,}
  \hlkwc{title} \hlstd{=} \hlstr{"A Diagnostic to Assess the Fit of a Variogram to Spatial Data"}\hlstd{,}
  \hlkwc{author} \hlstd{=} \hlstr{"Ronald Barry"}\hlstd{,} \hlkwc{journaltitle} \hlstd{=} \hlstr{"Journal of Statistical Software"}\hlstd{,}
                 \hlkwc{volume} \hlstd{=} \hlnum{1}\hlstd{,} \hlkwc{number} \hlstd{=} \hlnum{1}\hlstd{)}
\hlstd{bib}
\end{alltt}
\begin{verbatim}
## [1] R. Barry. "A Diagnostic to Assess the Fit of a Variogram to
## Spatial Data". In: _Journal of Statistical Software_ 1.1 (Aug.
## 1996).
\end{verbatim}
\begin{alltt}
\hlkwd{toBiblatex}\hlstd{(bib)}
\end{alltt}
\begin{verbatim}
## @Article{barry1996,
##   date = {1996-08},
##   title = {A Diagnostic to Assess the Fit of a Variogram to Spatial Data},
##   author = {Ronald Barry},
##   journaltitle = {Journal of Statistical Software},
##   volume = {1},
##   number = {1},
## }
\end{verbatim}
\end{kframe}
\end{knitrout}

\Biblatex{} offers a huge amount of additional functionality compared to \Bibtex{}.  For the full details, one can see the 253 page user manual \citep{biblatex}.  \Biblatex{} expands the number of automatically recognized entry types and fields offered by \Bibtex{}, allowing for much more detailed bibliographic entries,  while still maintaining compatibility with \Bibtex{}.  For example, to handle an \texttt{arXiv} eprint in \Bibtex{}, one needs to use or create a special \Bibtex{} style, or perhaps use the `note' and `year' fields in unintended ways.  The below entry is used in an attempt to cite a submitted manuscript of the first author's using the \code{bibentry} function. 

\begin{knitrout}
\definecolor{shadecolor}{rgb}{0.973, 0.973, 0.973}\color{fgcolor}\begin{kframe}
\begin{alltt}
\hlkwd{bibentry}\hlstd{(}\hlstr{"misc"}\hlstd{,} \hlkwc{key} \hlstd{=} \hlstr{"mclean2013bayesian"}\hlstd{,} \hlkwc{author} \hlstd{=} \hlstr{"M. W. McLean and 
         F. Scheipl and G. Hooker and S. Greven and D. Ruppert"}\hlstd{,}
         \hlkwc{title} \hlstd{=} \hlstr{"Bayesian Functional Generalized Additive Models 
                 with Sparsely Observed Covariates"}\hlstd{,} \hlkwc{year} \hlstd{=} \hlstr{"Submitted"}\hlstd{,}
         \hlkwc{note} \hlstd{=} \hlstr{"arXiv eprint: 1305.3585"}\hlstd{)}
\end{alltt}
\begin{verbatim}
## McLean MW, Scheipl F, Hooker G, Greven S and Ruppert D
## (Submitted). "Bayesian Functional Generalized Additive Models with
## Sparsely Observed Covariates." arXiv eprint: 1305.3585.
\end{verbatim}
\end{kframe}
\end{knitrout}

Though arXiv provides suggestions for creating \Bibtex{} entries for their papers (\url{http://arxiv.org/hypertex/bibstyles/}), there is a frustrating lack of consistency in how people choose to create \Bibtex{} entries for their arXiv papers.  In \Biblatex{}, there is greatly expanded support for electronic publications with fields for eprint, eprinttype, eprintclass, urldate, and pubstate.  One can cite the same article in \Biblatex{} without the need of a special \code{.bst} file or the note field using
 %<<bibexample, cache=FALSE, eval=FALSE, results='markup', tidy=FALSE>>=
 \begin{verbatim}
 @misc{mclean2013bayesian,
   author = {M. W. McLean and F. Scheipl and G. Hooker
                 and S. Greven and D. Ruppert},
   title = {Bayesian Functional Generalized Additive Models 
                 with Sparsely Observed Covariates},
   urldate = {2013-10-06},
   date = {2013},
   eprinttype = {arxiv},
   eprintclass = {stat.ME},
   eprint = {1305.3585},
   pubstate = {submitted},
 }
 \end{verbatim}
 The entry can be created using the \code{BibEntry} function in \ourpkg{}
\begin{knitrout}
\definecolor{shadecolor}{rgb}{0.973, 0.973, 0.973}\color{fgcolor}\begin{kframe}
\begin{alltt}
\hlkwd{BibEntry}\hlstd{(}\hlstr{"misc"}\hlstd{,} \hlkwc{key} \hlstd{=} \hlstr{"mclean2013bayesian"}\hlstd{,} \hlkwc{author} \hlstd{=} \hlstr{"McLean, M. W. and 
         Scheipl, F. and Hooker, G. and Greven, S. and Ruppert, D."}\hlstd{,}
         \hlkwc{title} \hlstd{=} \hlstr{"Bayesian Functional Generalized Additive Models 
                 with Sparsely Observed Covariates"}\hlstd{,} \hlkwc{urldate} \hlstd{=} \hlstr{"2013-10-06"}\hlstd{,}
         \hlkwc{date} \hlstd{=} \hlstr{"2013"}\hlstd{,} \hlkwc{eprinttype} \hlstd{=} \hlstr{"arxiv"}\hlstd{,} \hlkwc{eprintclass} \hlstd{=} \hlstr{"stat.ME"}\hlstd{,}
         \hlkwc{eprint} \hlstd{=} \hlstr{"1305.3585"}\hlstd{,} \hlkwc{pubstate} \hlstd{=} \hlstr{"submitted"}\hlstd{)}
\end{alltt}
\begin{verbatim}
## [1] McLean, M. W., Scheipl, et al. _Bayesian Functional
## Generalized Additive Models with Sparsely Observed Covariates_.
## 2013. arXiv: 1305.3585 [stat.ME]. (Visited on 10/06/2013).
## Submitted.
\end{verbatim}
\end{kframe}
\end{knitrout}

In \Biblatex{} the `eprint' identifier will automatically become a hyperlink to the paper on \texttt{arXiv}.  

The \code{bibentry} class supports \Bibtex{}-style crossreferencing, while the \code{BibEntry} class.  Cross references are handled specially when indexing and searching \code{BibEntry} objects and discussed in Section~\ref{searchsec}.  In a similar vain as cross-referencing, \Biblatex{} supports an entry type ``XData'' which is never printed, but may be used to store fields that are shared by several entries.  Entries can specify a field `xdata' containing a comma separated list of keys belonging to XData entries that the entry inherits from.  The following example demonstrates its use for online references available on arXiv, and uses the \code{c} operator for combining \code{BibEntry} objects.
\begin{knitrout}
\definecolor{shadecolor}{rgb}{0.973, 0.973, 0.973}\color{fgcolor}\begin{kframe}
\begin{alltt}
\hlstd{bib} \hlkwb{<-} \hlkwd{BibEntry}\hlstd{(}\hlkwc{bibtype}\hlstd{=}\hlstr{"XData"}\hlstd{,} \hlkwc{key} \hlstd{=} \hlstr{"statME"}\hlstd{,} \hlkwc{eprinttype} \hlstd{=} \hlstr{"arxiv"}\hlstd{,}
                \hlkwc{eprintclass} \hlstd{=} \hlstr{"stat.ME"}\hlstd{)}
\hlstd{bib} \hlkwb{<-} \hlkwd{c}\hlstd{(bib,} \hlkwd{BibEntry}\hlstd{(}\hlkwc{bibtype}\hlstd{=}\hlstr{"XData"}\hlstd{,} \hlkwc{key} \hlstd{=} \hlstr{"online2013"}\hlstd{,} \hlkwc{year} \hlstd{=} \hlstr{"2013"}\hlstd{,}
                       \hlkwc{urldate} \hlstd{=} \hlstr{"2013-12-20"}\hlstd{))}
\hlkwd{toBiblatex}\hlstd{(bib)}
\end{alltt}
\begin{verbatim}
## @XData{statME,
##   eprinttype = {arxiv},
##   eprintclass = {stat.ME},
## }
## 
## @XData{online2013,
##   year = {2013},
##   urldate = {2013-12-20},
## }
\end{verbatim}
\begin{alltt}
\hlstd{bib} \hlkwb{<-} \hlkwd{c}\hlstd{(bib,} \hlkwd{BibEntry}\hlstd{(}\hlkwc{bibtype}\hlstd{=}\hlstr{"Online"}\hlstd{,} \hlkwc{key}\hlstd{=}\hlstr{"mclean2013rlrt"}\hlstd{,}
  \hlkwc{author} \hlstd{=} \hlstr{"Mathew McLean and Giles Hooker and David Ruppert"}\hlstd{,}
  \hlkwc{title} \hlstd{=} \hlstr{"Restricted Likelihood Ratio Tests for Scalar-on-Function Regression"}\hlstd{,}
  \hlkwc{eprint} \hlstd{=} \hlstr{"1310.5811"}\hlstd{,} \hlkwc{url} \hlstd{=} \hlstr{"http://arxiv.org/abs/1310.5811"}\hlstd{,}
  \hlkwc{xdata} \hlstd{=} \hlstr{"statME,online2013"}\hlstd{))}
\hlstd{bib} \hlkwb{<-} \hlkwd{c}\hlstd{(bib,} \hlkwd{BibEntry}\hlstd{(}\hlkwc{bibtype}\hlstd{=}\hlstr{"Online"}\hlstd{,} \hlkwc{key}\hlstd{=}\hlstr{"mclean2013bayesian"}\hlstd{,}
  \hlkwc{author} \hlstd{=} \hlkwd{paste}\hlstd{(}\hlstr{"Mathew McLean and Fabian Scheipl and Giles Hooker"}\hlstd{,}
                \hlstr{"and Sonja Greven and David Ruppert"}\hlstd{),}
  \hlkwc{title} \hlstd{=} \hlkwd{paste}\hlstd{(}\hlstr{"Bayesian Functional Generalized Additive Models"}\hlstd{,}
               \hlstr{"for Sparsely Observed Covariates"}\hlstd{),}
  \hlkwc{eprint} \hlstd{=} \hlstr{"1305.3585"}\hlstd{,} \hlkwc{url} \hlstd{=} \hlstr{"http://arxiv.org/abs/1305.3585"}\hlstd{,}
  \hlkwc{xdata} \hlstd{=} \hlstr{"statME,online2013"}\hlstd{))}
\hlstd{bib}
\end{alltt}
\begin{verbatim}
## XData: online2013
## 
## XData: statME
## 
## [1] M. McLean, G. Hooker and D. Ruppert. _Restricted Likelihood
## Ratio Tests for Scalar-on-Function Regression_. 2013. arXiv:
## 1310.5811 [stat.ME]. <URL: http://arxiv.org/abs/1310.5811>
## (visited on 12/20/2013).
## 
## [2] M. McLean, F. Scheipl, G. Hooker, et al. _Bayesian Functional
## Generalized Additive Models for Sparsely Observed Covariates_.
## 2013. arXiv: 1305.3585 [stat.ME]. <URL:
## http://arxiv.org/abs/1305.3585> (visited on 12/20/2013).
\end{verbatim}
\end{kframe}
\end{knitrout}

The cross-referencing system in \Biblatex{} and \ourpkg{} is more sophisticated than the symmetric field mapping system used in \Bibtex{}, allowing for less cluttering and duplication of fields.  In \Biblatex{}, the ``InBook'' entry type is used for a self-contained work with its own title within a book, as opposed to simply referring to an untitled part of a book as in \Bibtex{}. In the following example, involving an `InBook' entry inheriting from a `Book' entry, there is no need to create a `booktitle' field duplicating the `title' field in the parent entry to pass on to the child entry, and there is also no need to create an empty `subtitle' field in the child entry to ensure it does not incorrectly inherit the `subtitle` of the parent.  
\begin{knitrout}
\definecolor{shadecolor}{rgb}{0.973, 0.973, 0.973}\color{fgcolor}\begin{kframe}
\begin{alltt}
\hlkwd{c}\hlstd{(}\hlkwd{BibEntry}\hlstd{(}\hlstr{"book"}\hlstd{,} \hlkwc{key} \hlstd{=} \hlstr{"parent"}\hlstd{,} \hlkwc{title} \hlstd{=} \hlstr{"The Book Title"}\hlstd{,} \hlkwc{year} \hlstd{=} \hlnum{2012}\hlstd{,}
           \hlkwc{subtitle} \hlstd{=} \hlstr{"The Book Subtitle"}\hlstd{,} \hlkwc{author} \hlstd{=} \hlstr{"Book Author"}\hlstd{,}
           \hlkwc{publisher} \hlstd{=} \hlstr{"A publisher"}\hlstd{),} \hlkwd{BibEntry}\hlstd{(}\hlstr{"inbook"}\hlstd{,} \hlkwc{key} \hlstd{=} \hlstr{"child"}\hlstd{,}
           \hlkwc{crossref} \hlstd{=} \hlstr{"parent"}\hlstd{,} \hlkwc{title} \hlstd{=} \hlstr{"The Title of the In Book Entry"}\hlstd{,}
           \hlkwc{author} \hlstd{=} \hlstr{"In Book Author"}\hlstd{))}
\end{alltt}
\begin{verbatim}
## [1] B. Author. _The Book Title. The Book Subtitle_. A publisher,
## 2012.
## 
## [2] I. B. Author. "The Title of the In Book Entry". In: B. Author.
## _The Book Title. The Book Subtitle_. A publisher, 2012.
\end{verbatim}
\end{kframe}
\end{knitrout}

\ourpkg{} recognizes some, but not all, localization keys defined by default in \Biblatex{}.  A localization key is a special value that \Biblatex{} parses for certain fields and replaces with predefined text called the `localization string` when printing the bibliography.  In the example below I use localization keys to specify the roles of editors using the `editortype` field and refer to portions of a text using the `bookpagination` field.
\begin{knitrout}
\definecolor{shadecolor}{rgb}{0.973, 0.973, 0.973}\color{fgcolor}\begin{kframe}
\begin{alltt}
\hlkwd{BibEntry}\hlstd{(}\hlkwc{bibtype}\hlstd{=}\hlstr{"Collection"}\hlstd{,} \hlkwc{key} \hlstd{=} \hlstr{"jaffe"}\hlstd{,} \hlkwc{editor} \hlstd{=} \hlstr{"Phillip Jaff\textbackslash{}u00eb"}\hlstd{,}
  \hlkwc{title} \hlstd{=} \hlstr{"Regesta Pontificum Romanorum ab condita ecclesia ad annum post
  Christum natum \{MCXCVIII\}"}\hlstd{,} \hlkwc{date} \hlstd{=} \hlstr{"1885/1888"}\hlstd{,}
  \hlkwc{editora} \hlstd{=} \hlstr{"S. Loewenfeld and F. Kaltenbrunner and P. Ewald"}\hlstd{,}
  \hlkwc{editoratype} \hlstd{=} \hlstr{"redactor"}\hlstd{,} \hlkwc{totalpages} \hlstd{=} \hlstr{"10"}\hlstd{,} \hlkwc{bookpagination} \hlstd{=} \hlstr{"section"}\hlstd{)}
\end{alltt}
\begin{verbatim}
## [1] P. Jaffë, ed. _Regesta Pontificum Romanorum ab condita
## ecclesia ad annum post Christum natum MCXCVIII_. Red. by S.
## Loewenfeld, F. Kaltenbrunner and P. Ewald. 1885-1888.
\end{verbatim}
\end{kframe}
\end{knitrout}

\subsection[Reading .bib Files Into R]{Reading \code{.bib} Files Into \R{}} 
\ourpkg{} provides the function \code{ReadBib} for parsing \code{.bib} files in \Biblatex{} or \Bibtex{} format and creating \code{BibEntry} objects from them.  This function is based on the \code{read.bib} function in package \pkg{bibtex} \citep{Rbibtex}, which uses code for parsing \Bibtex{} files from \citet{bibparser}.  \code{ReadBib} expands on \code{read.bib} by providing \Biblatex{} support; by having an argument/option \code{check}, which can be disabled, and checks that each entry in the file has all the fields required by that \Biblatex{} or \Bibtex{} entry type; and also has expanded handling of name list fields to ensure that complicated names are correctly converted to \code{person} objects.  This last feature is important when searching the \code{BibEntry} object later; as it would not be possible to properly search by parts of a name (such as family name only) if a name has not been correctly converted to a \code{person} object.  Since it is often not necessary in \Biblatex{} to provide all the "required" fields for an entry, it can be useful to be able turn off the check for required fields in \R{} when one wants to work with entries that are missing some fields.  For example, the sample \code{.bib} file that comes with the \Biblatex{} package, and is also included with \ourpkg{} for demonstration purposes has three entries that are missing required fields.  The default behaviour is to not add these entries, but this can be changed.
\begin{knitrout}
\definecolor{shadecolor}{rgb}{0.973, 0.973, 0.973}\color{fgcolor}\begin{kframe}
\begin{alltt}
\hlstd{file} \hlkwb{<-} \hlkwd{system.file}\hlstd{(}\hlstr{"Bib"}\hlstd{,} \hlstr{"biblatexExamples.bib"}\hlstd{,} \hlkwc{package} \hlstd{=} \hlstr{"RefManageR"}\hlstd{)}
\hlstd{bib} \hlkwb{<-} \hlkwd{ReadBib}\hlstd{(file,} \hlkwc{check} \hlstd{=} \hlstr{"error"}\hlstd{)}
\end{alltt}
\begin{lstlisting}
## Ignoring entry titled "The Chicago Manual of Style" because A bibentry of bibtype 'Manual' has to specify the field: c("author", "editor")
## Ignoring entry titled "CTAN" because A bibentry of bibtype 'Online' has to specify the field: c("author", "editor")
## Ignoring entry titled "Computers and Graphics" because A bibentry of bibtype 'Periodical' has to specify the field: editor
\end{lstlisting}\begin{alltt}
\hlstd{bib} \hlkwb{<-} \hlkwd{ReadBib}\hlstd{(file,} \hlkwc{check} \hlstd{=} \hlnum{FALSE}\hlstd{)}
\hlkwd{print}\hlstd{(bib[}\hlkwd{c}\hlstd{(}\hlstr{"cms"}\hlstd{,} \hlstr{"jcg"}\hlstd{,} \hlstr{"ctan"}\hlstd{)],} \hlkwc{.opts} \hlstd{=} \hlkwd{list}\hlstd{(}\hlkwc{check.entries} \hlstd{=} \hlnum{FALSE}\hlstd{))}
\end{alltt}
\begin{verbatim}
## [1] _The Chicago Manual of Style. The Essential Guide for Writers,
## Editors, and Publishers_. 15th ed. Chicago, Ill.: University of
## Chicago Press, 2003. ISBN: 0-226-10403-6.
## 
## [2] _Computers and Graphics_. 35.4 (2011): _Semantic 3D Media and
## Content_. ISSN: 0097-8493.
## 
## [3] _CTAN. The Comprehensive TeX Archive Network_. 2006. <URL:
## http://www.ctan.org> (visited on 10/01/2006).
\end{verbatim}
\end{kframe}
\end{knitrout}

\subsection{Creating Citations From PDFs}
Using the function \code{ReadPDFs} and the freely available software Poppler (\url{http://poppler.freedesktop.org}), it is possible to create references from PDFs stored on a user's machine.  The user specifies a directory containing PDFs (or a single PDF file) which are then read by Poppler and converted to \code{.txt} files which are read into \R{}, parsed into citations, and output as a \code{BibEntry} object.  

The function will first search the text for a Document Object Identifier (DOI), and if one is found, the citation information will be downloaded from CrossRef using their API.  This feature will be discussed in more detail in the next section.  The function also works especially well with PDFs downloaded from \url{jstor.org} by recognizing the format of the cover page that JSTOR generates.  This allows for detailed and accurate citations to be obtained.  The function also recognizes papers downloaded from \url{http://arXiv.org} and parses the arXiv identifier in its current and pre-March 2007 format.

If there is no DOI available and the document does not have a JSTOR cover page, it is considerably more difficult to obtain an accurate citation.  The function is often able to recover the title, author, and date information.  It can parse journal title, volume, and issue information if it is present in an obvious format.  Articles with complicated formatting and missing the features discussed in the previous paragraph are not likely to be parsed correctly and the user will have to manually edit the entries, which will be covered in a later section.
With the following code, Windows binaries of Poppler are downloaded along with some PDFs
to test out the function.
\begin{knitrout}
\definecolor{shadecolor}{rgb}{0.973, 0.973, 0.973}\color{fgcolor}\begin{kframe}
\begin{alltt}
\hlstd{tmpdir} \hlkwb{<-} \hlkwd{tempdir}\hlstd{()}
\hlstd{tmpfile} \hlkwb{<-} \hlkwd{tempfile}\hlstd{(}\hlstr{".zip"}\hlstd{, tmpdir)}
\hlkwd{download.file}\hlstd{(}\hlstr{"http://dl.dropbox.com/u/3291828/Poppler/poppler.0.22.0_win32.zip"}\hlstd{,}
              \hlstd{tmpfile)}
\hlkwd{unzip}\hlstd{(tmpfile,} \hlkwc{exdir} \hlstd{= tmpdir)}
\hlstd{curdir} \hlkwb{<-} \hlkwd{getwd}\hlstd{()}
\hlkwd{setwd}\hlstd{(}\hlkwd{file.path}\hlstd{(tmpdir,} \hlstr{"bin"}\hlstd{,} \hlkwc{fsep} \hlstd{=} \hlstr{"\textbackslash{}\textbackslash{}"}\hlstd{))}
\hlkwd{download.file}\hlstd{(}\hlstr{"http://www.jstatsoft.org/v56/i11/paper"}\hlstd{,} \hlstr{"jss.pdf"}\hlstd{,}
                \hlkwc{mode} \hlstd{=} \hlstr{"wb"}\hlstd{)}
\hlkwd{download.file}\hlstd{(}\hlstr{"http://arxiv.org/pdf/math/0703791"}\hlstd{,}
              \hlkwc{destfile} \hlstd{=} \hlstr{"FIZaop.pdf"}\hlstd{,} \hlkwc{mode} \hlstd{=} \hlstr{"wb"}\hlstd{)}
\hlkwd{download.file}\hlstd{(}\hlstr{"http://arxiv.org/pdf/math/0703858"}\hlstd{,}
              \hlkwc{destfile} \hlstd{=} \hlstr{"PBHTaos.pdf"}\hlstd{,} \hlkwc{mode} \hlstd{=} \hlstr{"wb"}\hlstd{)}
\hlkwd{download.file}\hlstd{(}\hlstr{"http://biomet.oxfordjournals.org/content/83/4/715.full.pdf"}\hlstd{,}
  \hlkwc{destfile} \hlstd{=} \hlstr{"ADVb.pdf"}\hlstd{,} \hlkwc{mode} \hlstd{=} \hlstr{"wb"}\hlstd{)}
\hlkwd{download.file}\hlstd{(}\hlstr{"http://www.jstor.org/stable/pdfplus/25645718.pdf"}\hlstd{,}
              \hlkwc{destfile} \hlstd{=} \hlstr{"jstor.pdf"}\hlstd{,} \hlkwc{mode} \hlstd{=} \hlstr{"wb"}\hlstd{)}
\end{alltt}
\end{kframe}
\end{knitrout}

\begin{knitrout}
\definecolor{shadecolor}{rgb}{0.973, 0.973, 0.973}\color{fgcolor}\begin{kframe}
\begin{alltt}
\hlstd{bib} \hlkwb{<-} \hlkwd{ReadPDFs}\hlstd{(}\hlstr{"."}\hlstd{)}
\end{alltt}
\begin{lstlisting}
## Getting Metadata for 5 pdfs...
## Getting 1 BibTeX entries from CrossRef...
## Done
\end{lstlisting}\begin{alltt}
\hlstd{bib}
\end{alltt}
\begin{verbatim}
## [1] A. AZZALINI and A. D. VALLE. "The Multivariate Skew-normal
## Distribution". In: _Biometrika_ (1996), pp. 715-726.
## 
## [2] R. Chepesiuk. "JSTOR and Electronic Archiving". In: _American
## Libraries_ 31.11 (2000), pp. 46-48. JSTOR: 25645718. <URL:
## http://www.jstor.org/stable/25645718>.
## 
## [3] S. Fang, P. Imkeller and T. Zhang. "Global flows for
## stochastic differential equations without global Lipschitz
## conditions". In: _The Annals of Probability_ 35.1 (Jan. 2007), pp.
## 180-205. DOI: 10.1214/009117906000000412. <URL:
## http://dx.doi.org/10.1214/009117906000000412>.
## 
## [4] S. Luo, Y. Chen, X. Su, et al. _Meta-analysis_. 2014.
## 
## [5] D. Paul, E. Bair, T. Hastie, et al. _“Pre-conditioning” For
## Feature Selection And Regression In High-dimensional Problems_.
## Apr. 16, 2013. arXiv: math/0703858v1. <URL:
## http://arxiv.org/abs/math/0703858v1>.
\end{verbatim}
\end{kframe}
\end{knitrout}

Note that entry [4] is not complete.  To clean up use \code{setwd(curdir)} and \code{unlink(tmpdir)}.

\subsection[Conversion of Other Object Types to Class BibEntry]{Conversion of Other Object Types to Class \code{BibEntry}}
The \code{as.BibEntry} function will convert objects of several other data types to \code{BibEntry} if they have the proper format.  Acceptable formats include a named \code{character vector} with entries for `bibtype', `key' and other fields for a single reference; a \code{list} of named \code{character vectors} for multiple references; \code{bibentry} objects; or a \code{data frame} with one row per entry, and columns for each field, including a column for `bibtype'.  For \code{data frames}, the row names should provide the `keys'.

The following example uses the function to create a \code{Bibentry} object containing entries for every installed \R{} package.  It makes use of the \code{installed.packages} and \code{citation} functions in the \pkg{utils} package.
\begin{knitrout}
\definecolor{shadecolor}{rgb}{0.973, 0.973, 0.973}\color{fgcolor}\begin{kframe}
\begin{alltt}
\hlstd{pkg.names} \hlkwb{<-} \hlkwd{rownames}\hlstd{(}\hlkwd{installed.packages}\hlstd{())}
\hlstd{pkg.bib} \hlkwb{<-} \hlkwd{lapply}\hlstd{(pkg.names,} \hlkwa{function}\hlstd{(}\hlkwc{pkg}\hlstd{)\{}
  \hlstd{refs} \hlkwb{<-} \hlkwd{as.BibEntry}\hlstd{(}\hlkwd{citation}\hlstd{(pkg))}
  \hlkwa{if} \hlstd{(}\hlkwd{length}\hlstd{(refs))}
    \hlkwd{names}\hlstd{(refs)} \hlkwb{<-} \hlkwd{make.unique}\hlstd{(}\hlkwd{rep}\hlstd{(pkg,} \hlkwd{length}\hlstd{(refs)))}
  \hlstd{refs}
\hlstd{\})}
\end{alltt}
\end{kframe}
\end{knitrout}

Keys may be extracted from \code{BibEntry} objects using either the \code{names} method or the \code{\bt$\bt} operator with \code{name} argument (the value to the right of the `\$' sign) equal to `key'.  The extra step involving the \code{\bt names<-\bt} method for \code{BibEntry} objects assigns a unique key to each entry, as the \code{citation} function does not provide a key for each entry and may return more than one reference for a single package.  At this point, because of the use of \code{lapply}, \code{pkg.bib} is a \code{list} of \code{BibEntry} objects, instead of a single \code{BibEntry} object.  One way to rectify this is using the internal function \code{MakeCitationList}.  Additionally, the \code{\bt names<-\bt} method can be used to assign keys.  Entries in our \code{BibEntry} object of packages may be referred to using the package name/key because of the special features of the \code{\bt[\bt} operator for \code{BibEntry} objects, which will be discussed in detail in a later section.
\begin{knitrout}
\definecolor{shadecolor}{rgb}{0.973, 0.973, 0.973}\color{fgcolor}\begin{kframe}
\begin{alltt}
\hlstd{pkg.bib} \hlkwb{<-} \hlstd{RefManageR:::}\hlkwd{MakeCitationList}\hlstd{(pkg.bib)}
\hlstd{pkg.bib[}\hlstr{"boot"}\hlstd{]}
\end{alltt}
\begin{verbatim}
## [1] A. Canty and B. D. Ripley. _boot: Bootstrap R (S-Plus)
## Functions_. R package version 1.3-9. 2013.
\end{verbatim}
\begin{alltt}
\hlstd{pkg.bib[}\hlkwc{key} \hlstd{=} \hlstr{"boot"}\hlstd{]}
\end{alltt}
\begin{verbatim}
## [1] A. Canty and B. D. Ripley. _boot: Bootstrap R (S-Plus)
## Functions_. R package version 1.3-9. 2013.
## 
## [2] A. C. Davison and D. V. Hinkley. _Bootstrap Methods and Their
## Applications_. ISBN 0-521-57391-2. Cambridge: Cambridge University
## Press, 1997. <URL: http://statwww.epfl.ch/davison/BMA/>.
\end{verbatim}
\end{kframe}
\end{knitrout}

Using \code{pkg.bib["boot"]} matches the entry with key exactly ``boot''.  Using \code{pkg.bib[key = "boot"]}, a (partial) match occurs for any entry whose `key' contains the \code{string} ``boot'', due to the default settings of the \code{\bt[\bt} operator discussed in Section~\ref{sec_manip}.
The \code{BibEntry} class also has methods \code{as.data.frame} and \code{unlist} to convert \code{BibEntry} objects to a data frame and unlist'ed vector, respectively.  The function \code{as.data.frame} will create  a \code{data frame} from a \code{BibEntry} object with each row corresponding to a unique entry and one column for every field present in the \code{BibEntry} object, including a column called `bibtype' for the type of entry.  \code{NA} values indicate that the field is not present in that entry (row of the \code{data frame}).  The row names will be the `key's of the entries.

\subsection{Setting Package Options}
The package contains a convenience function \code{BibOptions} for changing packages options.  This function behaves similarly to the \code{options} function in \pkg{base}.  This allows the user to set default values for several arguments to the most commonly used functions, so the user does not have to specify them each call.  Options may be specified in \code{name = value} pairs or as a \code{list}, and current values may be extracted by specifying a \code{character vector} of option names.  Discussion of most of the options is left to later sections of the document when introducing the functions the options affect.  One option already encountered, is whether to check to ensure that each entry has values for all the fields required by \Biblatex{} for its entry type.  As mentioned, though entries have required fields in \Biblatex{}, they are not really required as the package will work and generate a citation in any reasonable situation with missing fields.  The option is named \code{check.entries}, and the default value is \code{"error"}.  With this setting, an error is thrown when an attempt is made to use an entry with missing fields and a new entry is not created when an attempt is made to create an entry with missing fields.  The value \code{"warn"} results in a warning being thrown when an entry with missing fields is encountered, but execution will not be stopped.  Lastly, the value \code{FALSE} turns off checking of entries entirely.  In the following example, use of the \code{BibOptions} function is demonstrated and it is shown what happens for the different settings of the \code{check.entries} option.
\begin{knitrout}
\definecolor{shadecolor}{rgb}{0.973, 0.973, 0.973}\color{fgcolor}\begin{kframe}
\begin{alltt}
\hlkwd{BibOptions}\hlstd{(}\hlstr{"check.entries"}\hlstd{)}
\end{alltt}
\begin{verbatim}
## $check.entries
## [1] "error"
\end{verbatim}
\begin{alltt}
\hlkwd{BibEntry}\hlstd{(}\hlkwc{bibtype} \hlstd{=} \hlstr{"Online"}\hlstd{,} \hlkwc{key} \hlstd{=} \hlstr{"ctan"}\hlstd{,} \hlkwc{date} \hlstd{=} \hlstr{"2006"}\hlstd{,}
 \hlkwc{title} \hlstd{=} \hlstr{"The Comprehensive TeX Archive Network"}\hlstd{,} \hlkwc{url} \hlstd{=} \hlstr{"http://www.ctan.org"}\hlstd{)}
\end{alltt}
\begin{lstlisting}[style=error]
## Error: A bibentry of bibtype 'Online' has to specify the field: c("author", "editor")
\end{lstlisting}\begin{alltt}
\hlstd{old.opt.val} \hlkwb{<-} \hlkwd{BibOptions}\hlstd{(}\hlkwc{check.entries} \hlstd{=} \hlnum{FALSE}\hlstd{)}
\hlkwd{BibEntry}\hlstd{(}\hlkwc{bibtype} \hlstd{=} \hlstr{"Online"}\hlstd{,} \hlkwc{key} \hlstd{=} \hlstr{"ctan"}\hlstd{,} \hlkwc{date} \hlstd{=} \hlstr{"2006"}\hlstd{,}
 \hlkwc{title} \hlstd{=} \hlstr{"The Comprehensive TeX Archive Network"}\hlstd{,} \hlkwc{url} \hlstd{=} \hlstr{"http://www.ctan.org"}\hlstd{)}
\end{alltt}
\begin{verbatim}
## [1] _The Comprehensive TeX Archive Network_. 2006. <URL:
## http://www.ctan.org>.
\end{verbatim}
\begin{alltt}
\hlkwd{BibOptions}\hlstd{(old.opt.val)}  \hlcom{# restore the old value of the option}
\end{alltt}
\end{kframe}
\end{knitrout}

The default values of all options can be restored using \code{BibOptions(restore.defaults = TRUE)}.  A list of all options and their current values can be obtained by calling the function with no arguments, i.e. \code{BibOptions()}.
\section{Importing Citations From the Web}\label{sec_import}
\subsection{NCBI's Entrez}
The National Center for Biotechnology Information's Entrez Global Query Cross-Database Search provides access to a large number of databases related to health sciences. \ourpkg{} provides an interface to Entrez which allows for searching for references and parsing them to \code{BibEntry} objects.  Additionally, users may look up references given a set of PubMed ID's, search for ID's for references already stored in a \code{BibEntry} object, and search for related works to references already in \R{}.  The full documentation for Entrez is available in \citet{entrez}.

The first \ourpkg{} function to be discussed is \code{ReadPubMed}, which uses the ESearch E-Utility.  Among other features, ESearch is used to search any of Entrez's 38 databases (not just PubMed) using a query string and returns a list of entries in the database that match the query by their IDs. These IDs are then used to retrieve bibliographic information using another call to Entrez which is parsed into a \code{BibEntry} object and returned by \code{ReadPubMed}.  The next example does a simple search for some of Raymond J.\ Carroll's publications.
\begin{knitrout}
\definecolor{shadecolor}{rgb}{0.973, 0.973, 0.973}\color{fgcolor}\begin{kframe}
\begin{alltt}
\hlstd{rjc.pm} \hlkwb{<-} \hlkwd{ReadPubMed}\hlstd{(}\hlstr{"raymond j. carroll"}\hlstd{,} \hlkwc{database} \hlstd{=} \hlstr{"PubMed"}\hlstd{)}
\hlstd{rjc.pm[[}\hlnum{1L}\hlstd{]]}
\end{alltt}
\begin{verbatim}
## [1] Y. Cho, N. D. Turner, L. A. Davidson, et al. "Colon cancer
## cell apoptosis is induced by combined exposure to the n-3 fatty
## acid docosahexaenoic acid and butyrate through promoter
## methylation". In: _Experimental biology and medicine (Maywood,
## N.J.)_ (2014). DOI: 10.1177/1535370213514927. PMID: 24495951.
\end{verbatim}
\end{kframe}
\end{knitrout}

The \code{"..."} argument of \code{ReadPubMed} can be used to pass additional optional arguments to ESearch.  Among them are \code{retmax} to specify the maximum number of entries to return, \code{retstart} to specify the index of the first result to return, and \code{field} to search only a particular field of the entries for a match.  For controlling the date of the matches there are options, \code{datetype} which gives the type of date to consider when searching by date; for example \code{datetype = "pdat"} specifies to search by publication date and \code{datetype = "mdat"} specifies to search by modification date.  The \code{mindate} and \code{maxdate} options specify the minimum and maximum dates that the search results should be restricted to.  Dates should be in the format "YYYY", "YYYY/MM", or "YYYY/MM/DD".  Our next query returns one entry published in 2009 in the Journal of Statistical Software
\begin{knitrout}
\definecolor{shadecolor}{rgb}{0.973, 0.973, 0.973}\color{fgcolor}\begin{kframe}
\begin{alltt}
\hlkwd{ReadPubMed}\hlstd{(}\hlstr{"journal of statistical software"}\hlstd{,} \hlkwc{field} \hlstd{=} \hlstr{"journal"}\hlstd{,} \hlkwc{retmax} \hlstd{=} \hlnum{1}\hlstd{,}
    \hlkwc{mindate} \hlstd{=} \hlnum{2009}\hlstd{,} \hlkwc{maxdate} \hlstd{=} \hlnum{2009}\hlstd{)}
\end{alltt}
\begin{verbatim}
## [1] S. Holmes, A. Kapelner and P. P. Lee. "An Interactive Java
## Statistical Image Segmentation System: GemIdent". In: _Journal of
## statistical software_ 30.10 (2009). PMID: 21614138.
\end{verbatim}
\end{kframe}
\end{knitrout}

The \code{GetPubMedRelated} function uses the ELink E-Utility to find related articles to a set of articles or IDs.  Either a character vector of IDs or a \code{BibEntry} object containing entries with `eprinttype' field equal to "pubmed" and pubmed ID's stored in the `eprint' field (the format expected by \Biblatex{} and also returned by the \code{ReadPubMed} function) should be specified for the \code{id} argument.  ELink can perform in two distinct ways given a set of IDs, either search for related articles for each ID in the set separately, or use the entire set at once to find articles that are related to every article specified by the set of IDs.  The latter type of behaviour is requested in \code{GetPubMedRelated} by specifying \code{batch.mode = TRUE} as an argument in the call.  In the below example I find related entries to the articles returned by the previous query for publications by RJC.
\begin{knitrout}
\definecolor{shadecolor}{rgb}{0.973, 0.973, 0.973}\color{fgcolor}\begin{kframe}
\begin{alltt}
\hlkwd{GetPubMedRelated}\hlstd{(rjc.pm,} \hlkwc{batch.mode} \hlstd{=} \hlnum{TRUE}\hlstd{,} \hlkwc{max.results} \hlstd{=} \hlnum{1}\hlstd{)}
\end{alltt}
\begin{verbatim}
## [1] J. Fan and Y. Wu. "Semiparametric estimation of covariance
## matrices for longitudinal data". In: _Journal of the American
## Statistical Association_ 103.484 (2008), pp. 1520-1533. DOI:
## 10.1198/016214508000000742. PMID: 19180247.
\end{verbatim}
\end{kframe}
\end{knitrout}

Entrez returns a similarity score with each returned citation giving a measure of how similar the returned entry is to the specified IDs.  These scores can be returned in the outputted \code{BibEntry} object in a field called `score' by specifying \code{return.sim.scores = TRUE} in the call.  Additionally, the IDs in the call that were used to determine the relation can be included in the output in a field called `PMIDrelated' if the argument \code{return.related.ids} is \code{TRUE}.  In the next example, \code{batch.mode = FALSE} is used and one related article is returned for each of two entries in \code{rjc.pm}.

\begin{knitrout}
\definecolor{shadecolor}{rgb}{0.973, 0.973, 0.973}\color{fgcolor}\begin{kframe}
\begin{alltt}
\hlkwd{BibOptions}\hlstd{(}\hlkwc{check.entries} \hlstd{=} \hlnum{FALSE}\hlstd{)}
\hlstd{ids} \hlkwb{<-} \hlstd{rjc.pm}\hlopt{$}\hlstd{eprint[}\hlnum{3}\hlopt{:}\hlnum{4}\hlstd{]}
\hlstd{ids}
\end{alltt}
\begin{lstlisting}[showstringspaces=false,style=output,columns=fullflexible,breaklines=true,inputencoding=utf8,extendedchars=	rue,breakautoindent=false,breakindent=0pt,inputencoding=utf8]
## $guenther2014healthy
## [1] "24453128"
## 
## $li2013selecting
## [1] "24376287"
\end{lstlisting}\begin{alltt}
\hlstd{related} \hlkwb{<-} \hlkwd{GetPubMedRelated}\hlstd{(ids,} \hlkwc{batch.mode} \hlstd{=} \hlnum{FALSE}\hlstd{,} \hlkwc{max.results} \hlstd{=} \hlkwd{c}\hlstd{(}\hlnum{1}\hlstd{,} \hlnum{1}\hlstd{),}
                      \hlkwc{return.sim.scores} \hlstd{=} \hlnum{TRUE}\hlstd{,} \hlkwc{return.related.ids} \hlstd{=} \hlnum{TRUE}\hlstd{)}
\hlkwd{toBiblatex}\hlstd{(related)}
\end{alltt}
\begin{lstlisting}[showstringspaces=false,style=output,columns=fullflexible,breaklines=true,inputencoding=utf8,extendedchars=	rue,breakautoindent=false,breakindent=0pt,inputencoding=utf8]
## @Article{guenther2008evaluation,
##   title = {Evaluation of the Healthy Eating Index-2005},
##   author = {Patricia M Guenther and Jill Reedy and Susan M Krebs-Smith and Bryce B Reeve},
##   year = {2008},
##   journal = {Journal of the American Dietetic Association},
##   volume = {108},
##   number = {11},
##   pages = {1854-64},
##   eprint = {18954575},
##   doi = {10.1016/j.jada.2008.08.011},
##   eprinttype = {pubmed},
##   score = {54583749},
##   pmidrelated = {24453128},
## }
## 
## @Article{seghouane2007criterion,
##   title = {The AIC criterion and symmetrizing the Kullback-Leibler divergence},
##   author = {Abd-Krim Seghouane and Shun-Ichi Amari},
##   year = {2007},
##   journal = {IEEE transactions on neural networks / a publication of the IEEE Neural Networks Council},
##   volume = {18},
##   number = {1},
##   pages = {97-106},
##   eprint = {17278464},
##   doi = {10.1109/TNN.2006.882813},
##   eprinttype = {pubmed},
##   score = {20610997},
##   pmidrelated = {24376287},
## }
\end{lstlisting}\end{kframe}
\end{knitrout}

The \code{LookupPubMedID} function is provided by \ourpkg{} to use Entrez's Ecitmatch to search for PubMed IDs for entries in an existing \code{BibEntry} object.  In the following, I read in a \Bibtex{} file of references to RJC papers from Google Scholar and search for PubMed ID's for the first ten entries.  If the search is successful and an ID is found, the corresponding entry is updated so that the `eprinttype' field is assigned the value “pubmed” and the `eprint' field is assigned the ID.
\begin{knitrout}
\definecolor{shadecolor}{rgb}{0.973, 0.973, 0.973}\color{fgcolor}\begin{kframe}
\begin{alltt}
\hlstd{file.name} \hlkwb{<-} \hlkwd{system.file}\hlstd{(}\hlstr{"Bib"}\hlstd{,} \hlstr{"RJC.bib"}\hlstd{,} \hlkwc{package} \hlstd{=} \hlstr{"RefManageR"}\hlstd{)}
\hlstd{bib} \hlkwb{<-} \hlkwd{ReadBib}\hlstd{(file.name)}
\hlstd{bib} \hlkwb{<-} \hlkwd{LookupPubMedID}\hlstd{(bib,} \hlkwd{seq_len}\hlstd{(}\hlnum{10}\hlstd{))}
\end{alltt}
\begin{lstlisting}
## Success for entries: 1, 2, 8, 9
\end{lstlisting}\begin{alltt}
\hlstd{bib[}\hlkwc{eprinttype} \hlstd{=} \hlstr{"pubmed"}\hlstd{][[}\hlnum{1L}\hlstd{]]}  \hlcom{# print entry for first located ID}
\end{alltt}
\begin{verbatim}
## [1] N. Serban, A. M. Staicu and R. J. Carroll. "Multilevel
## Cross-Dependent Binary Longitudinal Data". In: _Biometrics_ 69.4
## (2013), pp. 903-913. PMID: 24131242.
\end{verbatim}
\end{kframe}
\end{knitrout}

Finally, the \code{GetPubMedByID} function uses Entrez's Efetch to obtain bibliography data given a vector of PubMed ID's.  The just obtained PubMed IDs can be used to get the \Bibtex{} entry from Entrez and compare it with the one already in our bibliography from Google Scholar.
\begin{knitrout}
\definecolor{shadecolor}{rgb}{0.973, 0.973, 0.973}\color{fgcolor}\begin{kframe}
\begin{alltt}
\hlkwd{GetPubMedByID}\hlstd{(}\hlkwd{unlist}\hlstd{(bib}\hlopt{$}\hlstd{eprint)[}\hlnum{1L}\hlstd{])}
\end{alltt}
\begin{verbatim}
## [1] N. Serban, A. Staicu and R. J. Carroll. "Multilevel
## cross-dependent binary longitudinal data". In: _Biometrics_ 69.4
## (2013), pp. 903-13. DOI: 10.1111/biom.12083. PMID: 24131242.
\end{verbatim}
\end{kframe}
\end{knitrout}

If one wishes to use other NCBI E-Utilities and does not wish to work with \code{BibEntry} or \code{bibentry} objects, see the \pkg{rentrez} package \citep{rentrez}.
\subsection{Zotero}
\code{Zotero} is free, open source software for collecting and sharing bibliographic information.  \code{Zotero} can automatically retrieve bibliographic metadata that has been embedded in webpages using \code{ContextObjects in Spans} (\code{COinS}), and is thus a very convenient way to collect bibliographic information when browsing, for example, journal websites.  The \ourpkg{} package contains functions for querying existing \code{Zotero} libraries and converting the results to a \code{BibEntry} object and also for uploaded an existing \code{BibEntry} object to a Zotero library.  To use the Zotero API, one needs a Zotero account, a userID and an API key for the library one wishes to access.  The userID and API key for personal libraries may be found by logging in and visiting the page \url{https://www.zotero.org/settings/keys}.  The following call to \code{ReadZotero} searches for the first two references with the word `Bayesian' in the title contained in the library specified by the \code{`key'} parameter.
\begin{knitrout}
\definecolor{shadecolor}{rgb}{0.973, 0.973, 0.973}\color{fgcolor}\begin{kframe}
\begin{alltt}
\hlkwd{ReadZotero}\hlstd{(}\hlkwc{user} \hlstd{=} \hlstr{'1648676'}\hlstd{,} \hlkwc{.params} \hlstd{=} \hlkwd{list}\hlstd{(}\hlkwc{q} \hlstd{=} \hlstr{'bayesian'}\hlstd{,}
                               \hlkwc{key} \hlstd{=} \hlstr{'7lhgvcwVq60CDi7E68FyE3br'}\hlstd{,} \hlkwc{limit} \hlstd{=} \hlnum{2}\hlstd{))}
\end{alltt}
\begin{verbatim}
## [1] P. Müller and R. Mitra. "Bayesian Nonparametric Inference –
## Why and How". In: _Bayesian Analysis_ 8.2 (Jun. 2013), pp.
## 269-302. ISSN: 1936-0975. DOI: 10.1214/13-BA811. <URL:
## http://projecteuclid.org/euclid.ba/1369407550> (visited on
## 10/24/2013).
## 
## [2] K. Sriram, R. Ramamoorthi and P. Ghosh. "Posterior Consistency
## of Bayesian Quantile Regression Based on the Misspecified
## Asymmetric Laplace Density". In: _Bayesian Analysis_ 8.2 (Jun.
## 2013), pp. 479-504. ISSN: 1936-0975. DOI: 10.1214/13-BA817. <URL:
## http://projecteuclid.org/euclid.ba/1369407561> (visited on
## 10/24/2013).
\end{verbatim}
\end{kframe}
\end{knitrout}

\subsection{Google Scholar}
A function is provided for downloading citations from a public Google Scholar profile.  This function is partially based on the function \code{get_publications} in the \pkg{scholar} package \citep{scholar}, but provides additional functionality and processes the results into a \code{BibEntry} object.  The function requires the Google Scholar ID of the researcher of interest.  A user can obtain this ID by navigating to the researcher's Google Scholar profile and copying the value of the \code{user} parameter in the URL.  The profile must be public for the function to work.  The function assumes that each entry is either of type `Article' or type `Book'.  If any numbers are available with the entry relating to journal volume, number, or pages; then the entry will be classified as type `Article'.  Otherwise, the type will be `Book'.  The code that follows will return the Raymond J.\ Carroll's three most recent papers indexed by Google Scholar.
\begin{knitrout}
\definecolor{shadecolor}{rgb}{0.973, 0.973, 0.973}\color{fgcolor}\begin{kframe}
\begin{alltt}
\hlcom{## RJC's Google Scholar profile is at: }
\hlcom{## http://scholar.google.com/citations?user=CJOHNoQAAAAJ}
\hlstd{rjc.bib} \hlkwb{<-} \hlkwd{ReadGS}\hlstd{(}\hlkwc{scholar.id} \hlstd{=} \hlstr{'CJOHNoQAAAAJ'}\hlstd{,} \hlkwc{sort.by.date} \hlstd{=} \hlnum{TRUE}\hlstd{,}
                  \hlkwc{limit} \hlstd{=} \hlnum{3}\hlstd{)}
\hlstd{rjc.bib}
\end{alltt}
\begin{verbatim}
## [1] Y. Cho, N. D. Turner, L. A. Davidson, et al. "Colon cancer
## cell apoptosis is induced by combined exposure to the n-3 fatty
## acid docosahexaenoic acid and butyrate through promoter
## methylation". In: _Experimental Biology and Medicine_
## 1535370213514927 (2014).
## 
## [2] P. M. Guenther, S. I. Kirkpatrick, J. Reedy, et al. "The
## Healthy Eating Index-2010 Is a Valid and Reliable Measure of Diet
## Quality According to the 2010 Dietary Guidelines for Americans".
## In: _The Journal of nutrition, jn._ 113 (2014).
## 
## [3] M. P. Little, A. G. Kukush, S. V. Masiuk, et al. "Impact of
## Uncertainties in Exposure Assessment on Estimates of Thyroid
## Cancer Risk among Ukrainian Children and Adolescents Exposed from
## the Chernobyl Accident". In: _PLOS ONE_ 9.1 (2014).
\end{verbatim}
\end{kframe}
\end{knitrout}

The function also stores the number of citations of each result.  Each \code{BibEntry} will store the number of citations in a field \code{'cites'}, which is ignored when generating a bibliography by \Biblatex{} or \Bibtex{} without additional effort to handle a custom entry field.  The following code will obtain the second author's three most cited works according to Google Scholar and prints the citation count and entry type for each entry.
\begin{knitrout}
\definecolor{shadecolor}{rgb}{0.973, 0.973, 0.973}\color{fgcolor}\begin{kframe}
\begin{alltt}
\hlcom{## RJC's Google Scholar profile is at: }
\hlcom{## http://scholar.google.com/citations?user=CJOHNoQAAAAJ}
\hlstd{rjc.bib} \hlkwb{<-} \hlkwd{ReadGS}\hlstd{(}\hlkwc{scholar.id} \hlstd{=} \hlstr{'CJOHNoQAAAAJ'}\hlstd{,} \hlkwc{sort.by.date} \hlstd{=} \hlnum{FALSE}\hlstd{,}
                  \hlkwc{limit} \hlstd{=} \hlnum{3}\hlstd{)}
\hlstd{rjc.bib}
\end{alltt}
\begin{verbatim}
## [1] R. J. Carroll and D. Ruppert. _Transformation and weighting in
## regression_. CRC Press, 1988.
## 
## [2] R. J. Carroll, D. Ruppert, L. A. Stefanski, et al.
## _Measurement error in nonlinear models: a modern perspective_. CRC
## press, 2012.
## 
## [3] D. Ruppert, M. P. Wand and R. J. Carroll. _Semiparametric
## regression_. Cambridge University Press, 2003.
\end{verbatim}
\begin{alltt}
\hlkwd{cbind}\hlstd{(rjc.bib}\hlopt{$}\hlstd{cites, rjc.bib}\hlopt{$}\hlstd{bibtype)}
\end{alltt}
\begin{verbatim}
##                           [,1]   [,2]  
## carroll2012measurement    "2495" "Book"
## ruppert2003semiparametric "1931" "Book"
## carroll1988transformation "1416" "Book"
\end{verbatim}
\end{kframe}
\end{knitrout}

A shortcoming of this approach, is that long author lists, long titles, or long journal/publisher info can all lead to incomplete information being returned for those fields for the offending entries.  In this case, the \code{ReadGS} function will either not include entry or provide a add the entry with a warning depending on the value of the \code{check.entries} argument.
\begin{knitrout}
\definecolor{shadecolor}{rgb}{0.973, 0.973, 0.973}\color{fgcolor}\begin{kframe}
\begin{alltt}
\hlcom{## RJC's Google Scholar profile is at: }
\hlcom{## http://scholar.google.com/citations?user=CJOHNoQAAAAJ}
\hlstd{rjc.bib} \hlkwb{<-} \hlkwd{ReadGS}\hlstd{(}\hlkwc{scholar.id} \hlstd{=} \hlstr{'CJOHNoQAAAAJ'}\hlstd{,} \hlkwc{sort.by.date} \hlstd{=} \hlnum{FALSE}\hlstd{,}
                  \hlkwc{limit} \hlstd{=} \hlnum{10}\hlstd{,} \hlkwc{check.entries} \hlstd{=} \hlstr{'error'}\hlstd{)}
\end{alltt}
\begin{lstlisting}
## Incomplete author information for entry "Structure of dietary measurement error: results of the OPEN biomarker study" it will NOT be added
\end{lstlisting}\begin{alltt}
\hlstd{rjc.bib2} \hlkwb{<-} \hlkwd{ReadGS}\hlstd{(}\hlkwc{scholar.id} \hlstd{=} \hlstr{'CJOHNoQAAAAJ'}\hlstd{,} \hlkwc{sort.by.date} \hlstd{=} \hlnum{FALSE}\hlstd{,}
                  \hlkwc{limit} \hlstd{=} \hlnum{10}\hlstd{,} \hlkwc{check.entries} \hlstd{=} \hlstr{'warn'}\hlstd{)}
\end{alltt}
\begin{lstlisting}
## Incomplete author information for entry "Structure of dietary measurement error: results of the OPEN biomarker study" adding anyway
\end{lstlisting}\begin{alltt}
\hlkwd{length}\hlstd{(rjc.bib)} \hlopt{==} \hlkwd{length}\hlstd{(rjc.bib2)}
\end{alltt}
\begin{verbatim}
## [1] FALSE
\end{verbatim}
\begin{alltt}
\hlcom{## the offending entry.  RJC is missing because list of authors was too long}
\hlkwd{print}\hlstd{(rjc.bib2[}\hlkwc{title}\hlstd{=}\hlstr{'dietary measurement error'}\hlstd{],}
      \hlkwc{.opts} \hlstd{=} \hlkwd{list}\hlstd{(}\hlkwc{max.names} \hlstd{=} \hlnum{99}\hlstd{,} \hlkwc{bib.style} \hlstd{=} \hlstr{'alphabetic'}\hlstd{))}
\end{alltt}
\begin{verbatim}
## [Kip+03] V. Kipnis, A. F. Subar, D. Midthune, L. S. Freedman, R.
## Ballard-Barbash and and. "Structure of dietary measurement error:
## results of the OPEN biomarker study". In: _American Journal of
## Epidemiology_ 158.1 (2003), pp. 14-21.
\end{verbatim}
\end{kframe}
\end{knitrout}

\subsection{CrossRef}
The function \code{ReadCrossRef} uses the CrossRef Metadata Search API (\url{http://search.crossref.org/help/api}) to import references based on a search of CrossRef's nearly 60 million records.  Given a search term and possibly a search year, the function receives \Bibtex{} entries as JSON objects using the \pkg{RJSONIO} package \citep{RJSONIO}, which are saved to a temporary file and then read back into \R{} using the \code{ReadBib} function to be returned as a \code{BibEntry} object.
\begin{knitrout}
\definecolor{shadecolor}{rgb}{0.973, 0.973, 0.973}\color{fgcolor}\begin{kframe}
\begin{alltt}
\hlkwd{ReadCrossRef}\hlstd{(}\hlkwc{query} \hlstd{=} \hlstr{'rj carroll measurement error'}\hlstd{,} \hlkwc{limit} \hlstd{=} \hlnum{3}\hlstd{,}
             \hlkwc{sort} \hlstd{=} \hlstr{"relevance"}\hlstd{,} \hlkwc{min.relevance} \hlstd{=} \hlnum{80}\hlstd{,} \hlkwc{verbose} \hlstd{=} \hlnum{FALSE}\hlstd{)}
\end{alltt}
\begin{verbatim}
## [1] R. J. Carroll, D. Ruppert and L. A. Stefanski. "Measurement
## Error in Nonlinear Models".  (1995). DOI:
## 10.1007/978-1-4899-4477-1. <URL:
## http://dx.doi.org/10.1007/978-1-4899-4477-1>.
## 
## [2] R. J. Carroll, D. Ruppert and L. A. Stefanski. "Response
## Variable Error". In: _Measurement Error in Nonlinear Models_
## (1995), p. 229–242. DOI: 10.1007/978-1-4899-4477-1_13. <URL:
## http://dx.doi.org/10.1007/978-1-4899-4477-1_13>.
## 
## [3] D. Ruppert, M. P. Wand and R. J. Carroll. "Measurement Error".
## In: _Semiparametric Regression_ (2003), p. 268–275. DOI:
## 10.1017/cbo9780511755453.017. <URL:
## http://dx.doi.org/10.1017/cbo9780511755453.017>.
\end{verbatim}
\end{kframe}
\end{knitrout}

Although false negatives are rare, the CrossRef Metadata Search can be prone to false positives.  For this reason, it is important to specify the \code{min.relevance} argument.  Each reference returned by CrossRef comes with a relevancy score which is CrossRef's determination of how likely the reference is to be a match for the supplied query.  The maximum possible value is 100, so for the most strict possible matching, specify \code{min.relevance = 100}.  If the argument \code{verbose} is \code{TRUE}, then a message is printed with the relevancy score and full citation for each reference with a relevancy score greater than \code{min.reference} in addition to returning the references in a \code{BibEntry} object.
% # <<ReadCR2, tidy=TRUE, highlight=TRUE>>=
% # bib <- ReadCrossRef(query = 'rj carroll data', limit = 3, sort = "relevance", 
% #              min.relevance = 50, verbose = TRUE)
% # @
\section{Sorting, Printing, Opening, and Outputting to File}\label{sec_print}
\subsection{Printing}
A number of \Biblatex{} bibliography styles are available in \ourpkg{} for formatting and displaying citations.  The styles currently implemented are ``numeric'' (the default), ``authortitle'', ``authoryear'', ``alphabetic'', and ``draft''.  The ``authoryear'' style always begins with the family name of the first author and follows the list of authors with the year of publication in parentheses.  The other four styles all use the same format, differing only in the label they print before each entry.  Style ``numeric'' prints the numeric index of each entry in the bibliography, style ``authortitle'' uses no label, style ``alphabetic'' creates a label using the family names of the authors and the last two digits of the publication year, and style ``draft'' uses the entry key as the label.

Entries may be printed as plain text, HTML, \Bibtex{} format, \Biblatex{} format, as \R{} code, Markdown, or as a mixture of \Bibtex{} and plain text commonly used for citations.  For an example of the ``authoryear'' style
\begin{knitrout}
\definecolor{shadecolor}{rgb}{0.973, 0.973, 0.973}\color{fgcolor}\begin{kframe}
\begin{alltt}
\hlstd{file.name} \hlkwb{<-} \hlkwd{system.file}\hlstd{(}\hlstr{"Bib"}\hlstd{,} \hlstr{"biblatexExamples.bib"}\hlstd{,} \hlkwc{package} \hlstd{=} \hlstr{"RefManageR"}\hlstd{)}
\hlstd{bib} \hlkwb{<-} \hlkwd{ReadBib}\hlstd{(file.name,} \hlkwc{check} \hlstd{=} \hlnum{FALSE}\hlstd{)}
\hlkwd{print}\hlstd{(bib[}\hlkwc{author} \hlstd{=} \hlstr{"Nietzsche"}\hlstd{],} \hlkwc{.opts} \hlstd{=} \hlkwd{list}\hlstd{(}\hlkwc{bib.style} \hlstd{=} \hlstr{"authoryear"}\hlstd{))}
\end{alltt}
\begin{verbatim}
## Nietzsche, F. (1988a). _Sämtliche Werke. Kritische
## Studienausgabe_. Ed. by G. Colli and M. Montinari. 2nd ed. Vol.
## 15. 15 vols. München and Berlin and New York: Deutscher
## Taschenbuch-Verlag and Walter de Gruyter.
## 
## —–— (1988b). _Sämtliche Werke. Kritische Studienausgabe_. Vol. 1.:
## _Die Geburt der Tragödie. Unzeitgemäße Betrachtungen I-IV.
## Nachgelassene Schriften 1870-1973_. Ed. by G. Colli and M.
## Montinari. 2nd ed. München and Berlin and New York: Deutscher
## Taschenbuch-Verlag and Walter de Gruyter.
## 
## —–— (1988c). "Unzeitgemässe Betrachtungen. Zweites Stück. Vom
## Nutzen und Nachtheil der Historie für das Leben". In: F.
## Nietzsche.  _Sämtliche Werke. Kritische Studienausgabe_. Vol. 1.:
## _Die Geburt der Tragödie. Unzeitgemäße Betrachtungen I-IV.
## Nachgelassene Schriften 1870-1973_. Ed. by G. Colli and M.
## Montinari. München and Berlin and New York: Deutscher
## Taschenbuch-Verlag and Walter de Gruyter, pp. 243-334.
\end{verbatim}
\end{kframe}
\end{knitrout}

The package has a number of options similar to those available in \Biblatex{}, including \code{dashed} to control the use of dashes for duplicate authors as in the above example, \code{max.names} to control the number of names in name list fields that will be printed before they are truncated with ``et al.'', and \code{first.inits} to control whether given names are truncated to first initials or full names are used.  These options can be set using the \code{BibOptions} function or passed as options to the \code{.opts} argument of the \code{print} method. There is also a package option, \code{no.print.fields} for supressing the printing of certain fields.
\begin{knitrout}
\definecolor{shadecolor}{rgb}{0.973, 0.973, 0.973}\color{fgcolor}\begin{kframe}
\begin{alltt}
\hlstd{old.opts} \hlkwb{<-} \hlkwd{BibOptions}\hlstd{(}\hlkwc{bib.style} \hlstd{=} \hlstr{"alphabetic"}\hlstd{,} \hlkwc{max.names} \hlstd{=} \hlnum{2}\hlstd{,}
                       \hlkwc{first.inits} \hlstd{=} \hlnum{FALSE}\hlstd{)}
\hlstd{bib[}\hlkwc{bibtype} \hlstd{=} \hlstr{"report"}\hlstd{]}
\end{alltt}
\begin{verbatim}
## [CC78] Willy W. Chiu and We Min Chow. _A Hybrid Hierarchical Model
## of a Multiple Virtual Storage (MVS) Operating System_. Research
## rep. RC-6947. IBM, 1978.
## 
## [PFT99] Jitendra Padhye, Victor Firoiu, et al. _A Stochastic Model
## of TCP Reno Congestion Avoidance and Control_. Tech. rep. 99-02.
## Amherst, Mass.: University of Massachusetts, 1999.
\end{verbatim}
\begin{alltt}
\hlkwd{BibOptions}\hlstd{(old.opts)}  \hlcom{# reset to original values}
\hlkwd{print}\hlstd{(bib[[}\hlnum{19}\hlstd{]],} \hlkwc{.opts} \hlstd{=} \hlkwd{list}\hlstd{(}\hlkwc{style} \hlstd{=} \hlstr{"html"}\hlstd{,} \hlkwc{no.print.fields} \hlstd{=} \hlstr{"url"}\hlstd{,}
      \hlkwc{bib.style} \hlstd{=} \hlstr{"authortitle"}\hlstd{))}
\end{alltt}
\begin{verbatim}
## <p><cite>Spiegelberg, H.
## &ldquo;&ldquo;Intention&rdquo; und &ldquo;Intentionalität&rdquo; in
## der Scholastik, bei Brentano und Husserl&rdquo;.
## In: <EM>Studia Philosophica</EM> 29 (1969), pp. 189-216.</cite></p>
\end{verbatim}
\end{kframe}
\end{knitrout}

The user can create a custom \Biblatex{} or \Bibtex{} bibliography style using the \code{bibstyle} fuction in the \code{tools} package.  To do this involves creating an environment containing functions for formatting entries of each type with signatures such as \code{formatArticle(paper)} and \code{formatBook(paper)}.

A downside of \Biblatex{} is that the majority of academic journals do not support its use, having long ago written a custom \code{bst} file for generating citations which can only be used by \Bibtex{}.  For this reason \ourpkg{} provides a \code{toBibtex} method returning a character vector with entries converted from \Biblatex{} to \Bibtex{} format.  Entries of a type that are not supported by \Bibtex{} will be converted to a type that is, e.g., entries of type `report' are converted to type `techreport'.  Other conversions include replacing the `date' field with a properly formatted `year' field (if year is not already present) and converting the `journaltitle' field to `journal'.  Since the cross-referencing system in \Bibtex{} is more limited than the one supported by \Biblatex{}, an attempt is made to ensure the cross-referencing will still work as expected in \Bibtex{}.  All fields not normally supported by \Bibtex{} are dropped unless they are specified in the argument \code{extra.fields}.  The argument \code{note.replace.field} can be used to specify fields to add to the `note' field in entries that are missing it.  As already demonstrated, the \code{toBiblatex} function will convert a \code{BibEntry} object to a character vector contains lines of the corresponding \Biblatex-formatted bibliography.  No fields are converted or dropped by this function; in this way it is very similar to the \code{toBibtex} method for \code{bibentry} objects.

\begin{knitrout}
\definecolor{shadecolor}{rgb}{0.973, 0.973, 0.973}\color{fgcolor}\begin{kframe}
\begin{alltt}
\hlstd{ref} \hlkwb{<-} \hlkwd{BibEntry}\hlstd{(}\hlstr{"thesis"}\hlstd{,} \hlkwc{key} \hlstd{=} \hlstr{"schieplthesis"}\hlstd{,} \hlkwc{date} \hlstd{=} \hlstr{"2011-03-17"}\hlstd{,} \hlkwc{url} \hlstd{=}
\hlstr{"http://edoc.ub.uni-muenchen.de/13028/"}\hlstd{,} \hlkwc{urldate} \hlstd{=} \hlstr{"2014-03-06"}\hlstd{,} \hlkwc{title} \hlstd{=}
\hlstr{"Bayesian Regularization and Model Choice for Structured Additive Regression"}\hlstd{,}
\hlkwc{type} \hlstd{=} \hlstr{"phdthesis"}\hlstd{,} \hlkwc{institution} \hlstd{=} \hlstr{"LMU Munich"}\hlstd{,} \hlkwc{author} \hlstd{=} \hlstr{"Fabian Scheipl"}\hlstd{)}
\hlkwd{toBiblatex}\hlstd{(ref)}
\end{alltt}
\begin{lstlisting}[showstringspaces=false,style=output,columns=fullflexible,breaklines=true,inputencoding=utf8,extendedchars=	rue,breakautoindent=false,breakindent=0pt,inputencoding=utf8]
## @Thesis{schieplthesis,
##   date = {2011-03-17},
##   url = {http://edoc.ub.uni-muenchen.de/13028/},
##   urldate = {2014-03-06},
##   title = {Bayesian Regularization and Model Choice for Structured Additive Regression},
##   type = {phdthesis},
##   institution = {LMU Munich},
##   author = {Fabian Scheipl},
## }
\end{lstlisting}\begin{alltt}
\hlkwd{toBibtex}\hlstd{(ref,} \hlkwc{note.replace.field} \hlstd{=} \hlstr{"urldate"}\hlstd{)}
\end{alltt}
\begin{lstlisting}[showstringspaces=false,style=output,columns=fullflexible,breaklines=true,inputencoding=utf8,extendedchars=	rue,breakautoindent=false,breakindent=0pt,inputencoding=utf8]
## @PhdThesis{schieplthesis,
##   url = {http://edoc.ub.uni-muenchen.de/13028/},
##   title = {Bayesian Regularization and Model Choice for Structured Additive Regression},
##   author = {Fabian Scheipl},
##   year = {2011},
##   month = {mar},
##   school = {LMU Munich},
##   note = {Last visited on 03/06/2014},
## }
\end{lstlisting}\end{kframe}
\end{knitrout}

The function \code{WriteBib}, based on the function \code{write.bib} in the package \pkg{bibtex} \citep{Rbibtex}, is provided for writing a \code{BibEntry} object to a \code{bib} file in \Biblatex{} or \Bibtex{} format using \code{toBiblatex} and \code{toBibtex}, respectively, depending on the value of the \code{biblatex} logical argument to \code{WriteBib}.  In the next example I write the previous thesis reference to a file in \Bibtex{} format, and for demonstration purposes only, read back in the \code{.bib} file using \code{read.bib} in package \pkg{bibtex} so that a \code{bibentry} object is created instead of a \code{BibEntry} one.
\begin{knitrout}
\definecolor{shadecolor}{rgb}{0.973, 0.973, 0.973}\color{fgcolor}\begin{kframe}
\begin{alltt}
\hlstd{tmpfile} \hlkwb{<-} \hlkwd{tempfile}\hlstd{(}\hlkwc{fileext} \hlstd{=} \hlstr{".bib"}\hlstd{)}
\hlkwd{WriteBib}\hlstd{(ref,} \hlkwc{file} \hlstd{= tmpfile,} \hlkwc{biblatex} \hlstd{=} \hlnum{FALSE}\hlstd{,} \hlkwc{verbose} \hlstd{=} \hlnum{FALSE}\hlstd{)}
\hlkwd{library}\hlstd{(bibtex)}
\hlkwd{read.bib}\hlstd{(tmpfile)}
\end{alltt}
\begin{verbatim}
## Scheipl F (2011). _Bayesian Regularization and Model Choice for
## Structured Additive Regression_. PhD thesis, LMU Munich. Last
## visited on 03/06/2014, <URL:
## http://edoc.ub.uni-muenchen.de/13028/>.
\end{verbatim}
\begin{alltt}
\hlkwd{unlink}\hlstd{(tmpfile)}
\end{alltt}
\end{kframe}
\end{knitrout}

\subsection{Sorting}
Nine different methods are available for sorting citations stored in a \code{BibEntry} object, corresponding to the ones predefined in \Biblatex{}.  Depending on the \code{bib.style} option, the default sorting method is \code{``nty''} to sort by `name' (`n'), then `title' (`t'), then `year'/`date' (`y').  Other possibilities are ``debug'' to sort by `key', ``none'' for no sorting, ``nyt'', ``nyvt'', ``anyt'', ``anyvt'', ``ynt'', and ``ydnt''; where the `a' stands for sorting by alphabetic label, `v' stands for sorting by `volume', and `yd' for sorting by `year'/`date' in descending order.

All sorting methods first consider the field `presort', if available. Entries with no `presort' field are assigned `presort' value ``mm''. Next the `sortkey' field is used.  When sorting by name, the `sortname' field is used first. If it is not present, the `author' field is used, if that is not present `editor' is used, and if that is not present `translator' is used.  When sorting by `title', first `sorttitle' is considered. Similarly, when sorting by `year', `sortyear' is first considered.  When sorting by `volume', if the field is present, it is padded to four digits with leading zeros; otherwise, the string ``0000'' is used.  When sorting by alphabetic label, first `shorthand' is considered, then `label', then `shortauthor', `shorteditor', `author', `editor', and `translator'. Refer to \citet[Sections~3.1.2.1 and 3.5 and Appendix~C.2][]{biblatex} for further details.

\subsection{Opening Connections to References}\label{sec_open}
Using the \code{open} method for \code{BibEntry} objects, it is possible to open a connection to a copy of an entry in the bibliography.  This will work for entries that have a value for the `file' field (specifying the path to a local copy of the reference), the `doi' field, the `url' field, or the `eprint' field when the `eprinttype' field is equal to \code{"jstor"}, \code{"arxiv"}, or \code{"pubmed"}.  Which of those fields are used and in which order they are checked for can be specified using the \code{open.field} argument.  The viewer to use can be specified (as a path) using the \code{viewer} argument.  By default the value \code{getOptions("pdfviewer")} is used when opening values stored in the `file' field, and the value of \code{getOptions("browser")} is used to open values in the other fields.  Recalling our bibliography of installed packages, the `url' field can be used to open the resource for the \pkg{base} package.  Additionally, a path to the \R{} Language Definition manual can be given in the `file' field, so that that can be opened instead when requested.
\begin{knitrout}
\definecolor{shadecolor}{rgb}{0.973, 0.973, 0.973}\color{fgcolor}\begin{kframe}
\begin{alltt}
\hlkwd{open}\hlstd{(pkg.bib[[}\hlstr{"base"}\hlstd{]])}  \hlcom{# will use the 'url' field}
\hlstd{pkg.bib[[}\hlstr{"base"}\hlstd{]]}\hlopt{$}\hlstd{file} \hlkwb{<-} \hlkwd{file.path}\hlstd{(}\hlkwd{R.home}\hlstd{(}\hlstr{"doc/manual"}\hlstd{),} \hlstr{"R-lang.pdf"}\hlstd{)}
\hlkwd{open}\hlstd{(pkg.bib[[}\hlstr{"base"}\hlstd{]],} \hlkwc{open.field} \hlstd{=} \hlkwd{c}\hlstd{(}\hlstr{"file"}\hlstd{,} \hlstr{"url"}\hlstd{))}
\end{alltt}
\end{kframe}
\end{knitrout}

\section[Searching and Manipulating BibEntry Objects]{Searching and Manipulating \code{BibEntry} Objects}\label{sec_manip}
\subsection{Extraction Operators - Searching and Indexing}\label{searchsec}
The extraction operator \code{\bt[\bt}, has been defined for \code{BibEntry} objects to allow for easily searching a database of references saved in a \code{BibEntry} object.  A different interface providing the same functionality is the function \code{SearchBib}.  Search options can be changed by set variables in the BibOptions object or alternatively specified directly as arguments to the function \code{SearchBib}.  BibLaTeX date fields (`date', `year', `origdate', `urldate', `eventdate') and name lists (`author', `editor', `editora', `editorb', `editorc', `translator', `commentator', `annotator', `introduction', `foreword', `afterword', `bookauthor', and `holder') are handled specially as outlined below.  Other fields can be searched using either exact string matching or regular expressions, with or without ignoring case (controlled via options \code{use.regex} and \code{ignore.case}, respectively).

Indices and search terms can be specified in a number of ways.  Similar to the default extraction operator for list objects, a vector of numeric indices or logical values can be given.  Additionally, a character vector of `key' values can be specified.  To search by field, a query can be specified with comma delimited \code{field=search.term} pairs, with \code{search.term} potentially being a vector with length greater than one to match multiple terms for \code{field} (think ``OR'').  Each \code{field = search.term} pair will have to match to declare a match for that entry (think ``AND'').  Multiple queries (``OR'') can be handled (involving different fields) by enclosing each separate query preferably in a \code{list} or alternatively, \code{c}.  For example, $\text{list}(field_{11} = \mathbf{search.term_{11}},field_{12}=\mathbf{search.term_{12}}),\text{list}(field_{21}=\mathbf{search.term_{21}})$.  If \code{c} is used instead of \code{list}, then the search terms \emph{must} have length one.  Examples will be provided shortly after discussing the special handling of date and name fields.  An `!' at the beginning of a search term can be used to negate a match (obviously, this can also be done using regular expressions).

Valid values for date fields in \Biblatex{} have the form \code{yyyy}, \code{yyyy-mm}, \code{yyyy-mm-dd}, and can be intervals of the form \code{yyyy/yyyy}, \code{yyyy-mm/yyyy-mm}, \code{yyyy-mm-dd/yyyy-mm-dd}.  The second date can be omitted in the interval to allow for open-ended end dates, e.g. \code{yyyy/}.  When searching using a date field, the search string should have one of these formats.  Additionally, the search string can be an interval with no start date, e.g. \code{date = ``/1980-06''} to return all entries published before June, 1980.  The \pkg{lubridate} package \citep{lubridate} is used to compare date fields, dates specified as intervals are converted to class \code{Interval} and non-interval dates are converted to class \code{POSIXct}.  The format \code{yyyy-mm} \emph{is} currently supported despite not being supported in base \R{} or \pkg{lubridate}.  For compatibility with \Bibtex{}, \Biblatex{} and \ourpkg{} support the fields \code{year} and \code{month}, which are used if the \code{date} field is missing. Whether to ignore month and day values, if available, and only compare based on the year portion of the date field, is controlled by the option \code{match.date}, which supports two values ``exact'' or ``year.only''.

When searching name list fields, the search term is expected to have the same format as used in a \code{.bib} file, e.g., \code{``Doe, Jr., John and Jane \{Doe Smith\}''}.  Names can be matched based on family names only, by family name and given name initials, or by full name, depending on the value of the option \code{match.author}.  

Entries containing valid \code{crossref} and \code{xdata} fields are expanded prior to searching, so that when a match is found for a field and value that a child entry inherits from its parent, the result is both the parent and child being returned.  If a match is found in a child entry and not in the parent, only the child entry is returned, but the returned entry will contain any fields it inherits from its parent.  Any xdata entries that the child references will also be returned.  Examples follow.

\begin{knitrout}
\definecolor{shadecolor}{rgb}{0.973, 0.973, 0.973}\color{fgcolor}\begin{kframe}
\begin{alltt}
\hlstd{file.name} \hlkwb{<-} \hlkwd{system.file}\hlstd{(}\hlstr{"Bib"}\hlstd{,} \hlstr{"biblatexExamples.bib"}\hlstd{,} \hlkwc{package} \hlstd{=} \hlstr{"RefManageR"}\hlstd{)}
\hlstd{bib} \hlkwb{<-} \hlkwd{ReadBib}\hlstd{(file.name,} \hlkwc{check} \hlstd{=} \hlnum{FALSE}\hlstd{)}
\hlcom{# by default match.author = 'family.only' and ignore.case = TRUE inbook}
\hlcom{# entry inheriting editor field from parent}
\hlstd{bib[}\hlkwc{editor} \hlstd{=} \hlstr{"westfahl"}\hlstd{]}
\end{alltt}
\begin{verbatim}
## [1] G. Westfahl, ed. _Space and Beyond. The Frontier Theme in
## Science Fiction_. Westport, Conn. and London: Greenwood, 2000.
## 
## [2] G. Westfahl. "The True Frontier. Confronting and Avoiding the
## Realities of Space in American Science Fiction Films". In: _Space
## and Beyond. The Frontier Theme in Science Fiction_. Ed. by G.
## Westfahl. Westport, Conn. and London: Greenwood, 2000, pp. 55-65.
\end{verbatim}
\begin{alltt}
\hlcom{# no match with parent entry, the returned child has inherited fields}
\hlstd{bib[}\hlkwc{author} \hlstd{=} \hlstr{"westfahl"}\hlstd{]}
\end{alltt}
\begin{verbatim}
## [1] G. Westfahl. "The True Frontier. Confronting and Avoiding the
## Realities of Space in American Science Fiction Films". In: _Space
## and Beyond. The Frontier Theme in Science Fiction_. Ed. by G.
## Westfahl. Westport, Conn. and London: Greenwood, 2000, pp. 55-65.
\end{verbatim}
\end{kframe}
\end{knitrout}

\begin{knitrout}
\definecolor{shadecolor}{rgb}{0.973, 0.973, 0.973}\color{fgcolor}\begin{kframe}
\begin{alltt}
\hlcom{# Entries published in Zürich (in bib file Z\{\textbackslash{}'u\}ich) OR entries written by}
\hlcom{# Aristotle and published before 1930}
\hlstd{bib[}\hlkwd{list}\hlstd{(}\hlkwc{location} \hlstd{=} \hlstr{"Zürich"}\hlstd{),} \hlkwd{list}\hlstd{(}\hlkwc{author} \hlstd{=} \hlstr{"Aristotle"}\hlstd{,} \hlkwc{year} \hlstd{=} \hlstr{"/1930"}\hlstd{)]}
\end{alltt}
\begin{verbatim}
## [1] Aristotle. _De Anima_. Ed. by R. D. Hicks. Cambridge:
## Cambridge University Press, 1907.
## 
## [2] Aristotle. _Physics_. Trans.  by P. H. Wicksteed and F. M.
## Cornford. New York: G. P. Putnam, 1929.
## 
## [3] Aristotle. _The Rhetoric of Aristotle with a commentary by the
## late Edward Meredith Cope_. Ed. by E. M. Cope. With a comment. by
## E. M. Cope. Vol. 3. 3 vols. Cambridge University Press, 1877.
## 
## [4] Homer. _Die Ilias_. Trans.  by W. Schadewaldt. With an intro.
## by J. Latacz. 3rd ed. Düsseldorf and Zürich: Artemis \& Winkler,
## 2004.
\end{verbatim}
\begin{alltt}
\hlkwd{length}\hlstd{(bib[}\hlkwc{author} \hlstd{=} \hlstr{"!knuth"}\hlstd{])}
\end{alltt}
\begin{verbatim}
## [1] 85
\end{verbatim}
\end{kframe}
\end{knitrout}

The list extraction operator, \code{\bt[[\bt}, is used for extacting \code{BibEntry} objects by position (an integer) or the entry key (a string).  Unlike the default operator, a vector of indices may be given to extract more than one entry at a time.

As with \code{bibentry} objects, the \code{\bt$\bt} operator for \code{BibEntry} objects is used to return a list containing the value of a particular field for all entries, with a value of \code{NULL} returned for entries that do not have the specified field.  A list of all entry types or keys for the \code{BibEntry} object, \code{bib}, can be obtained using \code{bib$bibtype} and \code{bib$key}, respectively.
\subsection{Assignment Operators} 
List assignment, \code{\bt[[<-\bt} is used for replacing one entry in a \code{BibEntry} object with another.  The below example uses a bibliography of just under 500 works of Raymond J.\ Carroll indexed on Google Scholar.  It contains a number of errors, some of which are corrected below to help demonstrate the use of the package.  I make use of the \code{logical} option \code{return.ind} to have the search return a \code{numeric vector} of indices as opposed to a \code{BibEntry} object.
\begin{knitrout}
\definecolor{shadecolor}{rgb}{0.973, 0.973, 0.973}\color{fgcolor}\begin{kframe}
\begin{alltt}
\hlstd{file.name} \hlkwb{<-} \hlkwd{system.file}\hlstd{(}\hlstr{"Bib"}\hlstd{,} \hlstr{"RJC.bib"}\hlstd{,} \hlkwc{package} \hlstd{=} \hlstr{"RefManageR"}\hlstd{)}
\hlstd{bib} \hlkwb{<-} \hlkwd{ReadBib}\hlstd{(file.name)}
\hlcom{## length(bib)}
\hlkwd{length}\hlstd{(bib)} \hlopt{==} \hlkwd{length}\hlstd{(bib[}\hlkwc{author} \hlstd{=} \hlstr{"Carroll"}\hlstd{])}
\end{alltt}
\begin{verbatim}
## [1] FALSE
\end{verbatim}
\begin{alltt}
\hlcom{# which entries are missing RJC?}
\hlstd{ind} \hlkwb{<-} \hlkwd{SearchBib}\hlstd{(bib,} \hlkwc{author} \hlstd{=} \hlstr{"!Carroll"}\hlstd{,} \hlkwc{.opts} \hlstd{=} \hlkwd{list}\hlstd{(}\hlkwc{return.ind} \hlstd{=} \hlnum{TRUE}\hlstd{))}
\hlstd{bib[ind]}\hlopt{$}\hlstd{author}
\end{alltt}
\begin{verbatim}
## $z2010oracle
## [1] "J G M AR TI NE Z" "R J C AR RO LL"  
## 
## $caroll2006measurement
## [1] "R J Caroll"      "D Ruppert"       "L A Stefanski"   "C M Crainiceanu"
## 
## $ll1996measurement
## [1] "R J C AR RO LL"
## 
## $wu1989estimation
## [1] "M C Wu"     "K R Bailey"
## 
## $caroll1989covariance
## [1] "R J Caroll"
\end{verbatim}
\end{kframe}
\end{knitrout}

Clearly, one paper is incorrectly attributed to RJC and the other four have spelling errors.  We thus drop that entry and correct the spelling on the other four entries.
\begin{knitrout}
\definecolor{shadecolor}{rgb}{0.973, 0.973, 0.973}\color{fgcolor}\begin{kframe}
\begin{alltt}
\hlstd{bib} \hlkwb{<-} \hlstd{bib[}\hlopt{-}\hlstd{ind[}\hlnum{4L}\hlstd{]]}
\hlstd{bib[}\hlkwc{author}\hlstd{=}\hlstr{"!Carroll"}\hlstd{]}\hlopt{$}\hlstd{author} \hlkwb{<-} \hlkwd{c}\hlstd{(}\hlstr{"Martinez, J. G. and Carroll, R. J."}\hlstd{,}
 \hlstr{"Carroll, R. J. and Ruppert, D. and Stefanski, L. A. and Crainiceanu, C. M."}\hlstd{,}
 \hlstr{"Carroll, R. J."}\hlstd{,} \hlstr{"Carroll, R. J."}\hlstd{)}
\hlkwd{length}\hlstd{(bib)} \hlopt{==} \hlkwd{length}\hlstd{(bib[}\hlkwc{author}\hlstd{=}\hlstr{"Carroll"}\hlstd{])}
\end{alltt}
\begin{verbatim}
## [1] TRUE
\end{verbatim}
\end{kframe}
\end{knitrout}

I can update different fields of multiple entries using the operator \code{\bt[<-\bt} as follows.  
\begin{knitrout}
\definecolor{shadecolor}{rgb}{0.973, 0.973, 0.973}\color{fgcolor}\begin{kframe}
\begin{alltt}
\hlkwd{BibOptions}\hlstd{(}\hlkwc{sorting} \hlstd{=} \hlstr{"none"}\hlstd{,} \hlkwc{bib.style} \hlstd{=} \hlstr{"alphabetic"}\hlstd{)}
\hlstd{bib[}\hlkwd{seq_len}\hlstd{(}\hlnum{3}\hlstd{)]}
\end{alltt}
\begin{verbatim}
## [SSC13] N. Serban, A. M. Staicu and R. J. Carroll. "Multilevel
## Cross-Dependent Binary Longitudinal Data". In: _Biometrics_ 69.4
## (2013), pp. 903-913.
## 
## [Jen+13] E. M. Jennings, J. S. Morris, R. J. Carroll, et al.
## "Bayesian methods for expression-based integration of various
## types of genomics data". In: _EURASIP Journal on Bioinformatics
## and Systems Biology_ 2013.1 (2013), pp. 1-11.
## 
## [Gar+13] T. P. Garcia, S. Müller, R. J. Carroll, et al.
## "Identification of important regressor groups, subgroups and
## individuals via regularization methods: application to gut
## microbiome data". In: _Bioinformatics, btt_ 608 (2013).
\end{verbatim}
\begin{alltt}
\hlstd{bib[}\hlkwd{seq_len}\hlstd{(}\hlnum{3}\hlstd{)]} \hlkwb{<-} \hlkwd{list}\hlstd{(}\hlkwd{c}\hlstd{(}\hlkwc{date}\hlstd{=}\hlstr{"2013-12"}\hlstd{),} \hlcom{## add month to Serban et al.}
        \hlkwd{c}\hlstd{(}\hlkwc{url}\hlstd{=}\hlstr{"http://bsb.eurasipjournals.com/content/2013/1/13"}\hlstd{,}
          \hlkwc{urldate} \hlstd{=} \hlstr{"2014-02-02"}\hlstd{),} \hlcom{## add URL and urldate to Jennings et al.}
        \hlkwd{c}\hlstd{(}\hlkwc{doi}\hlstd{=}\hlstr{"10.1093/bioinformatics/btt608"}\hlstd{,}
          \hlkwc{journal} \hlstd{=} \hlstr{"Bioinformatics"}\hlstd{))} \hlcom{## add DOI and correct journal}
\hlstd{bib[}\hlkwd{seq_len}\hlstd{(}\hlnum{3}\hlstd{)]}
\end{alltt}
\begin{verbatim}
## [SSC13] N. Serban, A. M. Staicu and R. J. Carroll. "Multilevel
## Cross-Dependent Binary Longitudinal Data". In: _Biometrics_ 69.4
## (Dec. 2013), pp. 903-913.
## 
## [Jen+13] E. M. Jennings, J. S. Morris, R. J. Carroll, et al.
## "Bayesian methods for expression-based integration of various
## types of genomics data". In: _EURASIP Journal on Bioinformatics
## and Systems Biology_ 2013.1 (2013), pp. 1-11. <URL:
## http://bsb.eurasipjournals.com/content/2013/1/13> (visited on
## 02/02/2014).
## 
## [Gar+13] T. P. Garcia, S. Müller, R. J. Carroll, et al.
## "Identification of important regressor groups, subgroups and
## individuals via regularization methods: application to gut
## microbiome data". In: _Bioinformatics_ 608 (2013). DOI:
## 10.1093/bioinformatics/btt608.
\end{verbatim}
\end{kframe}
\end{knitrout}

Notice that I set \code{sorting = "none"} above.  Sorting of the entries is done by default when printing, and after sorting the print order of entries is unlikely to correspond to the index order in the \code{BibEntry} object.  

A \code{BibEntry} object may be used as the replacement value.  A field may be removed by specifying its value be
set to the empty string \code{''}.
\begin{knitrout}
\definecolor{shadecolor}{rgb}{0.973, 0.973, 0.973}\color{fgcolor}\begin{kframe}
\begin{alltt}
\hlstd{bib2} \hlkwb{<-} \hlstd{bib[}\hlkwd{seq_len}\hlstd{(}\hlnum{3}\hlstd{)]}
\hlstd{bib2[}\hlnum{2}\hlopt{:}\hlnum{3}\hlstd{]} \hlkwb{<-} \hlstd{bib[}\hlnum{5}\hlopt{:}\hlnum{6}\hlstd{]}
\hlcom{# Note the Sarkar et al. entry is arXiv preprint with incorrect journal field}
\hlstd{bib2}
\end{alltt}
\begin{verbatim}
## [SSC13] N. Serban, A. M. Staicu and R. J. Carroll. "Multilevel
## Cross-Dependent Binary Longitudinal Data". In: _Biometrics_ 69.4
## (Dec. 2013), pp. 903-913.
## 
## [TDC13] C. D. Tekwe, A. R. Dabney and R. J. Carroll. "Application
## of Survival Analysis Methodology to the Quantitative Analysis of
## LC-MS Proteomics Data". In: _AMINO ACIDS_ 45.3 (2013), pp.
## 609-609.
## 
## [Sar+13] A. Sarkar, D. Pati, B. K. Mallick, et al. "Adaptive
## Posterior Convergence Rates in Bayesian Density Deconvolution with
## Supersmooth Errors". In: _arXiv preprint arXiv:_ 1308 (2013).
\end{verbatim}
\begin{alltt}
\hlcom{# Change type, remove journal, correct arXiv information}
\hlstd{bib2[}\hlnum{3}\hlstd{]} \hlkwb{<-} \hlkwd{c}\hlstd{(}\hlkwc{journal}\hlstd{=}\hlstr{''}\hlstd{,} \hlkwc{eprinttype} \hlstd{=} \hlstr{"arxiv"}\hlstd{,} \hlkwc{eprint} \hlstd{=} \hlstr{"1308.5427"}\hlstd{,}
           \hlkwc{eprintclass} \hlstd{=} \hlstr{"math.ST"}\hlstd{,} \hlkwc{pubstate} \hlstd{=} \hlstr{"submitted"}\hlstd{,} \hlkwc{bibtype} \hlstd{=} \hlstr{"Misc"}\hlstd{)}
\hlstd{bib2[}\hlnum{3}\hlstd{]}
\end{alltt}
\begin{verbatim}
## [Sar+13] A. Sarkar, D. Pati, B. K. Mallick, et al. _Adaptive
## Posterior Convergence Rates in Bayesian Density Deconvolution with
## Supersmooth Errors_. 2013. arXiv: 1308.5427 [math.ST]. Submitted.
\end{verbatim}
\end{kframe}
\end{knitrout}

\subsection{Merging}
The combine function, \code{c}, is available for concatenating multiple \code{BibEntry} objects, and has been inherited from the \code{bibentry} class.  Of course, this does not perform any checking for duplicate entries.  For this, there is the \code{base} package generics \code{anyDuplicated}, \code{duplicated}, and \code{unique}, which check vectors for duplicate elements.  However, if \code{BibEntry} objects have been compiled from a number of different sources, these functions may be too strict, declaring entries distinct even if only one field has a small difference between the two entries.  For this reason, an additional operator \code{'+'} is supplied along with a wrapper function \code{merge}, that compares entries only based on the fields specified by the user.  Given \code{BibEntry} objects \code{bib1} and \code{bib2}, \code{bib1 + bib2} will return \code{bib1} appended with all entries of bib2 that have been determined not be duplicates of entries already in \code{bib1} by comparing all fields in \code{BibOptions()$merge.fields.to.check}, which can include \code{bibtype} and \code{key}.  The function also checks if there are any duplicate keys in the result, and will force them to be unique if duplicates are detected using \code{make.unique}. 

\section[Using RefManageR in Dynamic Documents]{Using \ourpkg{} in Dynamic Documents}\label{sec_cite}
One may print citations from a \code{BibEntry} object and generate a bibliography for all citations.  This is especially useful for inclusion in \code{RMarkdown} or \code{RHTML} documents.  In those two situations, which can be specified by setting \code{style = "markdown"} and \code{style = "html"}, respectively, in the \code{BibOptions} function, hyperlinks will also automatically be generated.  These hyperlinks will either link to an external copy of the reference (using the same mechanism as the \code{open} method discussed in Section~\ref{sec_open}), or point from the citation to the bibliography entry and vice versa.  This behaviour is controlled by the package option \code{hyperlink}.  Owing to the features of the \code{\bt[\bt} operator for \code{BibEntry} objects, there is no need to restrict to only citing entries using their keys.

In addition to the main function \code{Cite}, the functions \code{Citet}, \code{Citep}, \code{TextCite}, and \code{AutoCite} are provided for convenience and mimic the output of the corresponding \LaTeX{} commands in the \code{natbib} \citep{natbib} and \code{biblatex} \LaTeX{} packages.  Using the option \code{cite.style}, one can specify `alphabetic', `numeric', or `authoryear'  citations.  Additionally, for the `numeric' style, one may specify the option \code{super} to include the numeric citations as superscripts.  The punctuation to use for enclosing the citations, separating multiple citations, etc. can be set using the \code{bibpunct} option.  These options can be set in \code{BibOptions} or specified as a list to the \code{.opts} argument to any of the citation functions.  A function \code{NoCite} is provided to include a reference in the bibliography without citing it.  The bibliography is printed using the \code{PrintBibliography} function.  It is very similar to the \code{print} method for the \code{BibEntry} class, except that it will only print the references that have been cited with one of the citation functions.  A demonstration follows.  Note that we change around the \code{cite.style} and \code{bib.style} to show the different styles, but one would normally set these at the start of the document and usually keep them set to be equal so that the labels in the citations and bibliography match.
\begin{knitrout}
\definecolor{shadecolor}{rgb}{0.973, 0.973, 0.973}\color{fgcolor}\begin{kframe}
\begin{alltt}
\hlkwd{BibOptions}\hlstd{(}\hlkwc{check.entries} \hlstd{=} \hlnum{FALSE}\hlstd{,} \hlkwc{bib.style} \hlstd{=} \hlstr{"authoryear"}\hlstd{,} \hlkwc{style} \hlstd{=} \hlstr{"text"}\hlstd{)}
\hlstd{file.name} \hlkwb{<-} \hlkwd{system.file}\hlstd{(}\hlstr{"Bib"}\hlstd{,} \hlstr{"biblatexExamples.bib"}\hlstd{,} \hlkwc{package} \hlstd{=} \hlstr{"RefManageR"}\hlstd{)}
\hlstd{bib} \hlkwb{<-} \hlkwd{ReadBib}\hlstd{(file.name)}
\end{alltt}
\end{kframe}
\end{knitrout}

\code{Citet(bib, "loh")} produces Loh (1992), a "textual" citation using an entry key.  It is possible to cite in parentheses by `year' using \code{Citep(bib, year = "1899", .opts = list(cite.style = "alphabetic"))} [Wil99].  Next, three works by Averroes are cited \code{AutoCite(bib, author = "averroes", .opts = list(super = TRUE, cite.style = "numeric"))} $^{[1; 2; 3]}$.  There is some support for resolving ambiguous citations; consider \code{Citet(bib, author = "Baez")} Baez and Lauda (2004a); Baez and Lauda (2004b).  Finally, the bibliography is printed using \code{PrintBibliography}.
\begin{knitrout}
\definecolor{shadecolor}{rgb}{0.973, 0.973, 0.973}\color{fgcolor}\begin{kframe}
\begin{alltt}
\hlkwd{PrintBibliography}\hlstd{(bib,} \hlkwc{.opts} \hlstd{=} \hlkwd{list}\hlstd{(}\hlkwc{bib.style} \hlstd{=} \hlstr{"alphabetic"}\hlstd{))}
\end{alltt}
\begin{verbatim}
## [Ave69] Averroes. _Drei Abhandlungen über die Conjunction des
## separaten Intellects mit dem Menschen. Von Averroes (Vater und
## Sohn), aus dem Arabischen übersetzt von Samuel Ibn Tibbon_. Ed. by
## J. Hercz. Trans.  by J. Hercz. Berlin: S.~Hermann, 1869.
## 
## [Ave82] Averroes. _The Epistle on the Possibility of Conjunction
## with the Active Intellect by Ibn Rushd with the Commentary of
## Moses Narboni_. Ed. by K. P. Bland. Trans.  by K. P. Bland.
## Moreshet: Studies in Jewish History, Literature and Thought 7. New
## York: Jewish Theological Seminary of America, 1982.
## 
## [Ave92] Averroes. _Des Averroës Abhandlung: "Über die Möglichkeit
## der Conjunktion" oder "Über den materiellen Intellekt"_. Ed. by L.
## Hannes. Trans.  by L. Hannes. With annots. by L. Hannes. Halle an
## der Saale: C.~A. Kaemmerer, 1892.
## 
## [BL04a] J. C. Baez and A. D. Lauda. "Higher-Dimensional Algebra V:
## 2-Groups". Version 3. In: _Theory and Applications of Categories_
## 12 (2004), pp. 423-491. arXiv: math/0307200v3.
## 
## [BL04b] J. C. Baez and A. D. Lauda. _Higher-Dimensional Algebra V:
## 2-Groups_. Oct. 27, 2004. arXiv: math/0307200v3.
## 
## [Loh92] N. C. Loh. "High-Resolution Micromachined Interferometric
## Accelerometer". MA Thesis. Cambridge, Mass.: Massachusetts
## Institute of Technology, 1992.
## 
## [Wil99] O. Wilde. _The Importance of Being Earnest: A Trivial
## Comedy for Serious People_. English and American drama of the
## Nineteenth Century. Leonard Smithers and Company, 1899. Google
## Books: 4HIWAAAAYAAJ.
\end{verbatim}
\end{kframe}
\end{knitrout}

Typically, when using \code{knitr}, one would load \ourpkg{}, load the bibliography, and set package options in a chunk at the start of the document using option \code{include = FALSE} and then include citations and print the bibliography with options \code{echo = FALSE} and \code{results = "asis"}.  To see demonstrations of these functions use in \code{RMarkdown} and \code{RHTML} documents and the hyperlinking features, see the package \code{vignettes} as well as the examples at \code{?Cite}.

\section{Conclusion}\label{sec_conc}
The \ourpkg{} package provides \R{} with considerable extra resources for working with bibliographic data; alleviating much of the difficulty of managing references from several different sources.  Functions have been introduced for importing references from a number of online resources and additionally for conveniently editing entries and creating new ones.  By implementing many of the features of \Biblatex{}, several shortcomings of working with \Bibtex{} format are removed. Conversion between different formats, bibliography styles, and between \Biblatex{} and \Bibtex{} is made easy with the package.  The user is able to be less dependent on remembering entry keys when writing a document and is able to make complicated searches using a simple syntax with the \code{\bt[\bt} operator.  As more and more researchers become aware of the benefits of working with \code{Markdown}, the citation, hyperlinking, and printing capabilities of \ourpkg{} will be a useful tool.

Future work on \ourpkg{} will include allowing for additional citation and bibliography styles,  making it easier for users to define custom styles, and creating support for \code{Pandoc} (\url{http://johnmacfarlane.net/pandoc/}) style citations.  Additionally, more work may be needed to ensure that searching and merging can be done very quickly for extremely large bibliographies for certain applications.  I also wish to explore creating a revamped version of the \code{citEntry} function in package \pkg{utils} to allow package developers to include citations in \Biblatex{} format in their packages.
\section*{Acknowledgements}
The author was supported in part by a postdoctoral award from the Texas A\&M Institute for Applied Mathematics and Computational Science, and in part by a grant from the National Cancer Institute (R37-CA057030, R.\ J.\ Carroll, P.I.).  He would also like to thank R.\ J.\ Carroll for helpful comments on the manuscript and for having so many articles to reference in examples.
\bibliography{biblatex}
\end{document}